\documentclass[usenatbib,psfig]{mnras}

\usepackage{epsfig}
\usepackage[justification=centering]{caption}
\usepackage{natbib}
\usepackage{wrapfig}
\usepackage{float}
\usepackage{dirtytalk}
\usepackage{placeins}
\usepackage{threeparttable}

\begin{document}


\title[Shear velocities and star formation]{Coupling local to global star formation in spiral galaxies: the effect of differential rotation}
\author[Aouad, James, \& Chilingarian]{Charles J. Aouad$^1$\thanks{E-mail:
charlesaouad@aascid.ae} , Philip A. James$^1$, Igor V. Chilingarian$^{2,3}$\\
$^1$ Astrophysics Research Institute, Liverpool John Moores University, IC2, Liverpool Science Park, 146  Brownlow Hill, Liverpool L3 5RF, UK\\
$^2$ Center for Astrophysics | Harvard and Smithsonian, 60 Garden St. MS09, Cambridge, MA 02138, USA\\
$^3$ Sternberg Astronomical Institute, M. V. Lomonosov Moscow State University, 13 Universitetsky prospect, 119234 Moscow, Russia
}

\date{Accepted . Received ; in original form }

\pagerange{\pageref{firstpage}--\pageref{lastpage}} \pubyear{2020}

\maketitle 

\label{firstpage}

\begin{abstract}
Star formation is one of the key factors that shapes galaxies. This process is relatively well understood from both simulations and observations on a small ``local’’ scale of individual giant molecular clouds and also on a ``global’’ galaxy-wide scale (e.g. the Kennicutt-Schmidt law). However, there is still no understanding on how to connect global to local star formation scales and whether this connection is at all possible. Here we analyze spatially resolved kinematics and the star formation rate density $\Sigma_{SFR} $ for a combined sample of 17 nearby spiral galaxies obtained using our own optical observations in H$\alpha$ for 9 galaxies and neutral hydrogen radio observations combined with a multi-wavelength spectral energy distribution analysis for 8 galaxies from the THINGS project. We show that the azimuthally averaged normalized star formation rate density in spiral galaxies on a scale of a few hundred parsecs is proportional to the kinetic energy of giant molecular cloud collisions due to differential rotation of the galactic disc. This energy is calculated from the rotation curve using the two Oort parameters $A$ and $B$ as $\log (\Sigma_{\mathrm{SFR}} / \mathrm{SFR}_{\mathrm{tot}}) \propto \log [2 A^2+ 5 B^2]$. The total kinetic energy of collision is defined by the shear velocity that is proportional to $A$ and the rotational energy of a cloud proportional to the vorticity $B$. Hence, shear does not act as a stabilizing factor for the cloud collapse thus reducing star formation but rather increases it by boosting the kinetic energy of collisions. This result can be a tool through which one can predict a radial distribution of star formation surface density using only a rotation curve.
\end{abstract}


\section{Introduction}
It is well established that stars form in the heart of dense cold molecular clouds  which are themselves embedded within the diffuse atomic gas.
However, although the presence of gas is a necessary condition for the process \citep{Kennicut,Kennicutt2007,Leroy2008,Bigiel2008,Wong2002,Schruba2011} it is not a sufficient one. In fact, star formation is very slow and inefficient \citep{Forbrich2009,Lada2010,Longmore2013,Kruijssen2014,Johnston2014,Barnes2017} even in cold dense regions of the interstellar medium (ISM) \citep{Querejeta2019}, and despite all investigations, whether from observations, numerical simulations and theoretical studies, there is still no consensus on how star formation proceeds, what factors regulate it, and how to connect local star formation regions on small scales to the global star formation properties on galactic scales. 

On the small scale (sub-parsec and few parsec scales) the process is a complex interplay between self gravity and  different opposing forces such as magnetic fields \citep{Boss2013,Dib2008,Dib2010}, thermal instabilities and radiative pressure \citep{Wada2000,Shadmehri2010,Kim2011}. The process is also controlled by other factors such as stellar feedback \citep{Murray2010,Dib2009}, turbulence \citep{Federratha2013}, chemical processes and kinematics \citep{Seigar2005,Dibetal2012,Meidt2013,Escala2008,Weidner2010,Hocuk2011,Colombo2018}.
 
On large scales, the relation is apparently simpler, as an empirical power law links the star formation rate surface density to the gas surface density (averaged on kpc scales) through $\Sigma_{SFR}\propto (\Sigma_{gas})^N$ \citep{Kennicut}. This relation is well known as the \say{Kennicutt-Schmidt law} (referred to throughout this paper as the KS law).
Although this empirical relation holds well when measured on kpc scales or on disk averaged data \citep{Bigiel2008,Leroy2008,Schruba2011}, it is not certain that it  holds on smaller scales. \citet{Onodera2010} found that the scatter of the KS relation increases on smaller scales until it breaks completely on the scale of molecular clouds (80 parsec scales). Additionally an increasing number of studies suggest that this relation varies between different galaxies, and even within different regions of the same galaxy \citep{Morokuma2017,Roman-Duval2016}.
\citet{HsiAn2014} find that the slope of the KS law increases from $\sim 1.4$ in areas of low $\Sigma_{H_2}$ to $\sim 2-3$ in regions of high $\Sigma_{H_2}$ in the same galaxy. They interpret this change as a shift in  regime in the sense that for regions of N $\sim 1.4$ the star formation is mainly triggered by gravitational instabilities while for the larger slope N=$\sim 2-3$ it is triggered by a combination of gravitational instabilities and cloud collisions.

The physical origin of the KS law is still debated, and there is some doubt that the overall large scale relation is the sum of similar ones over smaller scales. This has led many researchers to attempt to include parameters describing the kinematics of the galaxy, on small or large scales, as an additional ingredient to connect local star formation rates to global properties for a complete star formation theory \citep{Tan2000,krumholzmckee2005,krumholzmckee2009,Kennicutt1998}.  
\citet{Kennicutt1998} finds that  $\Sigma_{SFR}$ can be linked to the global galactic orbital angular frequency  $\Omega$ (taken at the outer radius used to perform the average) through $\Sigma_{SFR}\propto \Sigma_{gas} \Omega$, where $\Sigma_{gas}$ includes both the atomic and the  molecular components (H{\sc i} and H$_2$). 
\citet{silk1997} defines the relation between the $\Sigma_{SFR}$ and the $\Sigma_{gas}$ as $\Sigma_{SFR}=\epsilon\frac{\Sigma_{gas}}{\tau_{orb}}$ where $\tau_{orb}$ is the local orbital time and $\epsilon$ is the efficiency of star formation. 
Another recipe is $\Sigma_{SFR}\propto \Sigma_{gas}^N (\Omega-\Omega_p)$ \citep{wyse1986} where $\Omega_p$ is the pattern rotation frequency of the spiral arms. \citet{Tan2000} proposes a model in which the star formation rate increases in areas of strong shear mainly driven by cloud-cloud collisions due to the differential rotation of the disk.
 
It is also worth to mention that the classification of galaxies and therefore their kinematics on the global scale, seem to affect their global star formation rates \citep{Seigar2005,james2004,Lee2007} (i.e. the star formation activity increases when progressing along the Hubble sequence from early type to late type galaxies).

These findings suggest that the galaxy kinematics, in particular the way it rotates, play a role in regulating the star formation process on smaller scales. Even though the capacity of a galaxy to form stars is related to its total gas content, what sets the radial distribution of the star forming regions throughout the disk is still a subject of debate. 
The role of rotation has been investigated in several studies, both observational and through simulations, leading to conflicting conclusions \citep{Seigar2005,Dibetal2012,Colling2018,thilliez2014,Luna2005,thilliez2014,Colombo2018,Suwannajak2014,Escala2008,Weidner2010}.
A main challenge for researchers remains to find a star  formation theory that is consistent on all scales, and it becomes evident that the kinematics of the galaxy may be an important ingredient in such a theory.

This paper investigates the effect of the differential rotation on the star formation distribution throughout the disk, on scales of a few hundred parsecs. In Section 2 we describe the physics of differential rotation, in Section 3 and 4 we present our data and the data analysis methods, in Section 5 we show our results, leaving the discussion of these results to Section 6. In Section 7 we draw our conclusions and suggest some possible lines of future work.

\section{background}

Disk galaxies do not rotate as solid bodies, rather they rotate differentially. The consequence of this rotation is characterized  by two main physical effects: shear and vorticity, both of which are expressed as functions of the angular frequency $\Omega$ and the gradient of the angular frequency with respect to the radius $\frac{d\Omega}{dR}$ \citep{Binney2008}.

Shear is quantified by the first Oort parameter $A$, where
\begin{equation} 
A=-0.5R \frac{d\Omega}{dR}.
\label{eq1}
\end{equation}
Thus $A$ is the amount of stress between the differentially rotating annuli. It is a measure of how much a rotation curve deviates from a solid body trend in a specific position along the galactic radius, that is the angle between a straight line passing through the origin and the slope of the rotation curve at that particular point. It is a measure of the local shear values. The shearing motions will create a relative velocity between two points at different galactic radii and this relative velocity can be expressed as 
\begin{equation} 
  \dot{y} = b (\Omega-\frac{dV}{dR})=2\times b \times A 
  \label{eq2}
\end{equation}
 where $\dot{y}$ is the shear velocity (considering a frame of reference that is rotating at the same angular frequency $\Omega$ of the galaxy and $y$ is in the direction of rotation) and $b$ is the radial separation between two orbits. The increase of shear $A$ will increase the relative velocities between two points set at two different radii separated by a distance $b$.

Vorticity $\omega$ is the curl of the velocity vector field and is quantified by the second Oort parameter $B$ such that $\omega=\vert2B\vert$ \citep{binneymerrifield1998}
and 
\begin{equation} 
B=-(\Omega + 0.5R \frac{d\Omega}{dR}).
\end{equation}
Hence $B$ describes the tendency of the fluid to rotate about any given point. On the rotation curve, $B$ is proportional to the sum of two angles: the solid body rotation straight line passing through the origin and the derivative of the rotation curve at that particular point. In that sense it is the sum of both the galactic rotation and the additional differential rotation. In regions of solid body rotation it is exactly equal to the angular orbital galactic frequency. The differential rotation causes gravitationally bound parcels of fluid to rotate around their center, creating centrifugal forces that will counterbalance the collapsing effect of gravity. These forces will increase as the gas parcel shrinks while rotating faster due to conservation of angular momentum \citep{Toomre1964}.
For any gravitationally bound cloud, the radius at which the escape velocity is equal to the shear velocity is called the tidal radius and it represents the volume where the gas is confined gravitationally to the mass contained within its radius rather than escaping due to differential rotation. For clouds to not be shredded by shear, they must be contained within their tidal radius. Molecular clouds have been observed to have a radius in the same order as their tidal radius \citep{stark1978}. \citet{Gammie1991} use values of around 100 parsecs in areas of flat rotation; this same scale of the tidal radius has been used by several authors \citep{Tan2000,Tasker2009}.

 Furthermore, small radial perturbations caused by the non-uniform gravitational potential of the disk will cause periodic oscillations in the radial dimension, superimposed on the circular motion. From an inertial frame of reference e.g. that centred on the galaxy nucleus and rotating with the galaxy at any point $R$ along the galactic radius, these periodic perturbations will create additional elliptical motions, with an associated frequency $\kappa$ called the epicyclic frequency that can be measured from the rotation curve. $\kappa$ can be expressed as
\begin{equation}
\kappa=\sqrt[]{-4B\Omega}
\end{equation}
\citep{binneymerrifield1998}. \citet{Toomre1964} used $\kappa$  to quantify the stability of a differentially rotating disk in a theoretical frame. He argued that the disk can be stable against collapse on small scales because of thermal pressure (or velocity dispersion) and on larger scales because of differential rotation quantified by $\kappa$. The Toomre stability criterion Q is expressed as
\begin{equation}
Q= \frac{\kappa\sigma}{\pi G \Sigma}
\end{equation}
where $\sigma$ is the velocity dispersion, $G$ is the gravitational constant, $\kappa$ is the epicyclic frequency and $\Sigma$ is the gas surface density. If  $Q > 1$, this criterion is met, and the disk will be stable on all scales. 
The Toomre stability criterion can also be expressed with $B$ instead of $\kappa$ as both $B$ and $\kappa$ are proportional to the local rotational energy.

The Toomre stability criterion accurately predicts the star formation threshold in disk systems \citep{Martin2001}, however, it fails to do so for dwarf irregular galaxies \citep{Elson2012}. Furthermore, the presence of magnetic fields can transfer angular momentum away from the cloud \citep{Salmeron2009,Elmegreen1987}. Based on this reasoning, \citet{Elmegreen1993} and \citet{Hunter1998b} derive a stability parameter based on shear $A$, rather than epicyclic frequency $\kappa$. They argue that the competition between self gravity and shear is more relevant than the competition between self gravity and vorticity. This formalism has been adopted by many authors \citep{Dibetal2012,Luna2005,Takahira2018} and thus if gravitational perturbations are the main driver of star formation, then the limiting factor will be how quickly these perturbations grow before shear disrupts them and we should expect low star formation rates in areas of strong shear. However, if the star forming areas do not follow regions of strong shear then either shear is not enough to provide stability against collapse or another mechanism is driving the star formation.

 In a differentially rotating fluid it is possible that both shear and local rotation participate in setting the relative kinematics of the fluid parcels along the orbit, in the sense that gravitationally bound clouds can be rotating (due to vorticity as measured by $B$) while moving relative to each other at shear velocity (due to shear as measured by $A$).

In the current paper we use 2 sets of data to  investigate the relation between these 2 parameters (namely the Oort parameters $A$ and $B$), and the normalized star formation rate surface density distribution throughout the disk  (in azimuthally averaged radial rings) $\frac{\Sigma _{SFR}}{SFR_{total}} $ on scales of a few hundred parsecs.

\section{The Data}

The properties of the 17 spiral galaxies in the present study are listed in Table~\ref{tab:gals_obs}, including classification and distance properties which were taken from the NASA/IPAC Extragalactic Database (NED). Column 1 of Table~\ref{tab:gals_obs} contains the galaxy name; column 2 the galaxy classification; column 3 the adopted distance; column 4 the inclination angle; column 5 the position angle; column 6 the slit  position angle (where applicable); column 7 the total star formation rate; column 8 the star formation indicator and column 9 the method used to extract the rotation curve.

Our data consist of two sets referred here as: the optical galaxies (9 galaxies) from our own optical observations, and the radio galaxies (total 8)  which are galaxies with published kinematics and star formation rates from the THINGS \say{The H{\sc i} Nearby Galaxy Survey} \cite{walter2008}. 
Our choice of using two sets of data at different wavelengths and using different methods to extract our variables makes our findings more solid. For the optical data we use H$\alpha$ as tracer of star formation while for the radio data (taken from the literature) a combination of UV and mid-IR is used. In fact it is well known that recombination emission lines such as H$\alpha$ are not only due to star formation activity and they may be caused by other activities such as  shocks or AGNs, and additionally they can be significantly affected by extinction. The choice of using another set of data is motivated by this. Furthermore, the rotation curves of the optical dataset are derived using narrow band long slit spectroscopy, while for the radio dataset H{\sc i} integral field spectroscopy velocity maps are used, this also avoids the uncertainties of contamination due to gas local kinematics, velocity dispersion and turbulence of the H$_2$ regions.  

\subsection{The optical galaxy sample}

The optical data consist of $R$-band and H$\alpha$ narrow-band images. The images were taken in the 1m Jacobus Kapteyn Telescope  at the Observatorio del Roque de Los Muchachos on La Palma in the Canary Islands using the JKT CCD camera. The detector is SITe2, a device with 2048$\times$2048 24~$\mu$m pixels. The image scale is 0.33~arcsec~pixel$^{-1}$.
Long slit spectroscopic data were taken on the Isaac Newton Telescope (INT) using the Intermediate Dispersion Spectrograph (IDS) with a 235~mm focal length camera. The CCD used is a RED+2 with grating R1200Y. The CCD detector was used to maximise sensitivity in the region of H$\alpha$ emission, the most important feature for the present investigation. The slit width for all galaxy observations was 1.5$^{\prime\prime}$. The spatial pixel scale is 0.4$^{\prime\prime}$/pixel.
The unvignetted wavelength coverage was 1518\,\AA\ about a central wavelength of 4900\,\AA. The slit was not always aligned with the major axis of the galaxy being observed, requiring multiplicative geometric corrections to be made to the observed rotation curves, in addition to the usual inclination corrections.
 
\subsection{The radio galaxy sample (NGC)}
The THINGS \say{The H{\sc i} Nearby Galaxy Survey} \citep{walter2008} is a high spectral $>5.2$ km s$^{-1}$ and spatial (6~arcsec) resolution survey of H{\sc i} emission in 34 nearby galaxies obtained using the NRAO Very Large Array (VLA) at the National Radio Astronomy Observatory. The observed objects are at  distances between 2 and 15~Mpc with linear resolutions of $\sim$ 100 to 500 pc. The velocity field resolutions cover a range of 1.3 to 5.2 km~$s^{-1}$.

\begin{table*}
  \caption{Galaxy properties and observing details.
See Section 2 for descriptions of column entries.}
\begin{threeparttable}
  \begin{tabular}{llccccccc}
  \hline
 Name     &    Classn  & Dist$^a$ & i &P.A.  & Slit PA  &  SFR & SFRind$^b$ &Rotation curve$^c$\\
          &                & (Mpc) & (Deg) &  (Deg)     &  (Deg)     &  (${M_\odot}/yr$)  &           \\
  \hline
UGC\,5717  &   SABbc        &   29.043    &   61.1   &   16.5   & 189.99  &     0.9099($\pm$ 0.1677)   &  H$\alpha$ (J)  & H$\alpha$ long slit spect \\

UGC\,7315  &   SABbc     &   20.2   &   49.8   &   105     &       100    &  0.8649($\pm$ 0.2998)  & H$\alpha$ (J) & H$\alpha$ long slit spect\\

UGC\,2002 &   Sdm     &   11.684   &   23    &   70     &       89.98    & 0.1467($\pm$ 0.0834)  & H$\alpha$(J) & H$\alpha$ long slit spect\\

UGC\,5102   &   SABb     &   1797   &   69    &   131     &       0    &  1.7413($\pm$ 0.3253)   & H$\alpha$(J) & H$\alpha$ long slit spect\\

UGC\,4097  &   SAa   &   30.72   &   42   &   325.47     &      360.47    &  0.546($\pm$ 0.1701)   &  H$\alpha$(J) & H$\alpha$ long slit spect\\

UGC\,3580   &   SAa     &   21.902   &   62.2   &  2.8    &       29.99    &  0.5300 ($\pm$ 0.1301)   & H$\alpha$(J) & H$\alpha$ long slit spect\\

UGC\,2245   &   SABc     &  14.265   &   53   &   10.2   &       44.98    & 3.2074($\pm$ 0.8475)   & H$\alpha$(J) & H$\alpha$ long slit spect\\

UGC\,4779   &   SAc    &   22.477   &   61   &   87.3     &       219.9    &  1.5585($\pm$ 0.3438)    & H$\alpha$(J) & H$\alpha$ long slit spect\\

UGC\,3804   &   Scd     &   42.591   &   49.8   &  13.8     &       189.8    &  2.2694($\pm$ 0.4158)   & H$\alpha$(J) & H$\alpha$ long slit spect\\

NGC\,2403  &   SBc     &   3.2   &   63    &   124     &       -    &  0.382  & UV+NIR(L) & VF 1st moment (THINGS)dB\\

NGC\,2841  &   Sb     &   14.1   &   74    &   153    &       -   &  0.741   & UV+NIR(L) & VF 1st moment (THINGS)dB\\

NGC\,3521  &   SBbc    &   10.7   &   73    &   340     &       -    &  2.104  & UV+NIR(L) & tilted rings H{\sc i}(THINGS)dB\\

NGC\,3198  &   SBc     &   13.8   &   72    &   215     &       -    &  0.931   & UV+NIR(L) & tilted rings H{\sc i}(THINGS)dB\\

NGC\,6946  &   SBc    &   5.9   &   33    &   243     &       -    &  3.239  & UV+NIR(L) & VF 1st moment (THINGS)dB\\

NGC\,7793  &   Scd     &   3.9   &   50    &   290     &       -    &  0.235   & UV+NIR(L) & tilted rings H{\sc i}(THINGS)dB\\

NGC\,7331  &   SAb     &   14.7   &   76    &   168     &      -    &  2.987  & UV+NIR(L) &VF 1st moment (THINGS)dB\\

NGC\,5055  &   Sbc    &   10.1   &   59    &   102     &       -    &  2.123   & UV+NIR(L) & tilted rings H{\sc i}(THINGS)dB\\
\hline
\end{tabular}
\begin{tablenotes}
      \small
      \item[a:] Distances are taken from NED for the optical galaxies and from \citet{Leroy2008} for the radio galaxies.
      \item[b:] J: From \citet{james2004} for the optical galaxies. L: from \citet{Leroy2008} for the radio galaxies.
      \item[c:] Rotation curve data: dB = \citet{deblok2008}, using a tilted rings model.
      \item VF=cut through the velocity field of the first moment map from THINGS.
    \end{tablenotes}
\end{threeparttable}
\label{tab:gals_obs}
\end{table*}

\section{Data analysis}

\subsection{Building rotation curves and estimating shear $A$ and vorticity $B$}

For the optical data sample we constructed rotation curves using long-slit spectroscopy. The rotation velocity is calculated from the spectra binned to 3~pixels in the spatial direction, which corresponds a spatial sampling of 1.2~arcsec, similar to \citet{Deblok2002}. 
This corresponds to a physical sampling size in a range of 46~pc~bin$^{-1}$ for UGC~2002 to 184~pc~bin$^{-1}$ for UGC~4705 with a mean value of 59~pc~bin$^{-1}$.

We use the H$\alpha$ emission line as a velocity tracer. For every spatial bin along the slit, we extract ``slices''  using the {\sc figaro} command from the {\sc starlink} software package \citep{starlink}. 
An underlying local stellar continuum is fitted by a polynomial and subtracted from a spectrum, then a Gaussian function is fitted to the emission line. We build the rotation curve for both the receding and the approaching sides of a galaxy.
The centre of a galaxy was assumed to coincide with the maximum of the light on the slit. However, for some galaxies where this approach did not yield a reliable measure, we considered the systemic velocity to be that, which provides the smallest deviation between the two folded sides of the rotation curve.
Rotation velocities were corrected for inclination using the inclination angle computed from the axial ratio of a galaxy reported in the NED database. We also added an additional geometric correction in some galaxies, where the slits were not parallel to the major axis. 
For the subsequent analysis, we took a mean value between approaching and receding parts of each galaxy. The diversity of our rotation curves is similar to previous published rotation curves extracted using ionized gas emission lines \citep{swaters2000,Deblok2002,vogti2004}.

For the radio dataset we used the published rotation curves from \citet{deblok2008}, who used a tilted rings method to extract their rotation curves from atomic gas integral field spectroscopy data. The rotation curves of galaxies in the radio dataset extend to much further galactocentric radii than the optical sample.
For some galaxies (NGC~2403, NGC~2841, NGC~6946 and NGC~7331), the rotation curves for the inner part of the galaxy were not available, so we extracted them by cutting through the corresponding velocity fields (first moment maps).
The spatial sampling for radio rotation curves ranges from 97~pc~bin$^{-1}$ for NGC~6946 to 401~pc~bin$^{-1}$ for NGC~3198 with a mean value of 203~pc~bin$^{-1}$.

We fit a polynomial to the mean values measured (our polynomial range in degrees from 6 to 25 depending on each rotation curve) using a least square fitting technique. Our rotation curves are presented in the upper panels of Fig.\ref{fig1} and in  Appendix~A.
A polynomial fitting smooths a rotation curve by removing peaks connected to e.g. star-forming regions falling on the edge of the slit. It also makes it possible to use analytic derivatives instead of numerical ones for the subsequent analysis of shear and vorticity where $dV/dR$ are needed.
We calculate the absolute values of shear from the first Oort parameter $A$ using equation 2 and replacing $\Omega$ by $\frac{V}{R}$ we get  
\begin{equation} 
A=\vert-0.5(\frac{V}{R}-\frac{dV}{dR})\vert. 
\end{equation}

We calculate the absolute values of vorticity $B$ using equation~1. By replacing $\Omega$ by $\frac{V}{R}$, we get  
\begin{equation} 
B=\vert-0.5(\frac{V}{R}+\frac{dV}{dR})\vert
\end{equation}
where both $A$ and $B$ are expressed in km~s$^{-1}$~kpc$^{-1}$.

\subsection{Extracting Star Formation Rate Surface Density}

For the optical data, we use the narrow band H$\alpha$ images, to extract differential photometric measurements using elliptical apertures, since H$\alpha$ is a measure of the star formation rate \citep{Kennicutt1998,osterbrock2006,Davies2016}.
The measurements are taken differentially in elliptical annuli by subtracting the signal of each ellipse from the smaller ellipse starting from the centre and going towards the visible edge of a galaxy. The elliptical apertures are offset by 3 pixels, corresponding to a value of $0.993''$, consistent with the rotation curve sampling.

The eccentricity of the elliptical apertures is determined taking the $a/b$ ratio from NED, where \emph{a} and \emph{b} are the semi-major and semi-minor sizes of the galaxy respectively. The centre of each galaxy is defined from $R$ broad-band images.
We calibrate our photometric counts to star formation rate in ${M_\odot}$~yr$^{-1}~$kpc$^{-2}$ using total galaxy star formation rates from \citet{james2004}.
To calculate the surfaces we take the distances from NED and using the scale plate of the photometric data we convert to kpc$^{-2}$ correcting for inclination.
Our procedure results in azimuthally averaged star formation rates over the disk annuli.
Similar techniques have frequently been used in the past \citep[see e.g.][]{Martin2001,Wong2002,boissier2003}.

For the radio dataset we use the published spatially resolved star formation rates from \citet{Leroy2008}. Their star formation rate surface densities are extracted using combined data in far-ultraviolet from the GALEX nearby Galaxies survey \citep{giledepaz2007} and 24$\mu$m from the SINGS Spitzer Infrared Nearby Galaxy Survey. They argue that this combined tracer method has the advantage of being sensitive to both exposed and dust-embedded star formation.
The star formation rates are extracted in rings extending a few hundreds of parsecs (200~pc for NGC\,2403, NGC\,2841, NGC\,6946, and NGC\,7793; 700~pc for NGC\,7331; while NGC\,3184, NGC\,5055, and NGC\,3521 are sampled on scales of 400 to 600~pc.)

\subsection{Error Analysis}

For the optical sample, the errors on the rotation curve can be due to three main factors:
\begin{enumerate}
\item the error in determining the centre of the Gaussian fit: while for most of the points this error is a very small fraction of a pixel (we are dealing with very strong H$\alpha$ emissions), for some of them it is 0.1~pix and in rare cases it reaches 0.2~pix. We consider a mean error of 0.1~pix which corresponds to a value of 2.4 km~s$^{-1}$. This will yield different values for each galaxy after adding geometrical corrections for inclination and the slit misalignment with respect to the major axis.  
\item the error due to the random velocity of the ionized gas: we considered the random velocity to have a fixed value of 7~km~s$^{-1}$ across any galactic radius \citep[see e.g.][]{Gammie1991,Tan2000}
\item the error due to the folding of the two sides of a galaxy: since we are averaging our rotation curves to mean values calculated for the two sides, receding and approaching and the asymmetries in the folding are sometimes large. We take the error of the averaged velocity values to cover at least both values of the arms.
\end{enumerate}

We added all these errors in quadrature; the error on the rotation curve is propagated through the values of $A$ and $B$ by considering the error on $V$ and the error on the term $dV$ in radial spans consistent with our sampling.
The error on the star formation rate density are computed following  \citet{james2004} for the optical sample. We use the SFR density uncertainties reported by \cite{Leroy2008} for the radio sample.

\section{Results}

Our main results are presented in Fig.~\ref{fig1} and in Appendix \ref{appendixA}.
On the top panel we plot the rotation curve, and present separately the receding arm, the approaching arm, the mean values and the polynomial fit. 
The rotation curves do not follow a systematic trend of a ``rising'' and a ``flat'' part, rather their gradients keep changing, and our smoothed rotation curve polynomial fits  trace and describe these variations well. This has a remarkable consequence on the measurements of shear and vorticity. In a constant rising part of the rotation curve (a straight line passing through the origin) the shear value will always be zero and vorticity will always be equal to double the angular frequency (solid body rotation), while for the flat part, $A$ will decrease as $\propto 1/R$. 

However, this is not true for a rotation curve that is not simply a rising and a flat part. The slope of our polynomial fit keeps changing in both the ``rising'' and the ``flat'' parts. This is a key difference of our study with respect to most previous works, because most of them were done on kpc scales and ignored those small and real fluctuations of the rotation curve shape. 

The consequence of this can be seen in the values of $A$ and $B$ that we plot in the middle panel of Fig.\ref{fig1}. It is clear that the values of these two parameters do not follow a predicted pattern even in the innermost parts of the galaxy, because sudden changes in the slope of the curve cause large variations in the values of $A$ and $B$. It is also worth noting that $A$ increases in the parts of the rotation curve with negative gradients.

In the bottom panel of Fig.\ref{fig1}  we plot the value of the star formation rate fraction distribution $\Sigma_{\mathrm{SFR}} / \mathrm{SFR}_{\mathrm{tot}}$ that is the star formation rate surface density sampled on azimuthally averaged annuli normalized by the total star formation rate of the galaxy as a function of galactocentric radius. This represents a fraction of the total star formation rate of the galaxy in one particular ring. The normalization by the total star formation rate makes it possible to compare different galaxies with different gas contents, additionally this allows us to exclude central starbursts or nuclear star-forming regions induced by an inner bar. In Section~6 we point out that the combination $2A^2 + 5B^2$ measures a total kinetic energy of a rotating cloud moving with a shear velocity and rotating because of the vorticity originating from the differential rotation.
A visible trend can be seen, in which  $\Sigma_{\mathrm{SFR}} / \mathrm{SFR}_{\mathrm{tot}}$ follows the slope of $2A^2 + 5B^2 $.

This trend shows individually for all the galaxies and globally for 679 points when all values are combined together in the form of $log (\Sigma_{\mathrm{SFR}} / \mathrm{SFR}_{\mathrm{tot}}) \propto log [2 A^2+5 B^2]$. 
Figure~\ref{fig2} shows the radio data on the left panel and the optical data on the lower panel.
Figure~\ref{fig3} shows all galaxies combined together with the black line representing the  best fit.

We find a linear trend with a strong correlation coefficient of 0.836 for the subset shown in the white surface of our plot on Fig.~\ref{fig3}. We exclude the data points in light grey representing the low star formation rate at $\log (\Sigma_{\mathrm{SFR}} / \mathrm{SFR}_{\mathrm{tot}})<-4$ and the data points for $\log [2 A^2+5 B^2]> 4.2$ which correspond to very rapidly changing rotation curves, which might not be adequately described by our polynomial fitting procedure.

The best fitting relation to the data is linear. It has the form:
\begin{equation} 
 \log (\Sigma_{\mathrm{SFR}} / \mathrm{SFR}_{\mathrm{tot}})=-5.94_{[\pm0.11]}+1.01_{[\pm0.03]}\log(2A^2+5B^2).
 \label{eq8}
\end{equation}

\begin{figure*}
	\begin{tabular}{ll}
    \includegraphics[trim={0 0 0 0},clip,width=0.49\textwidth]{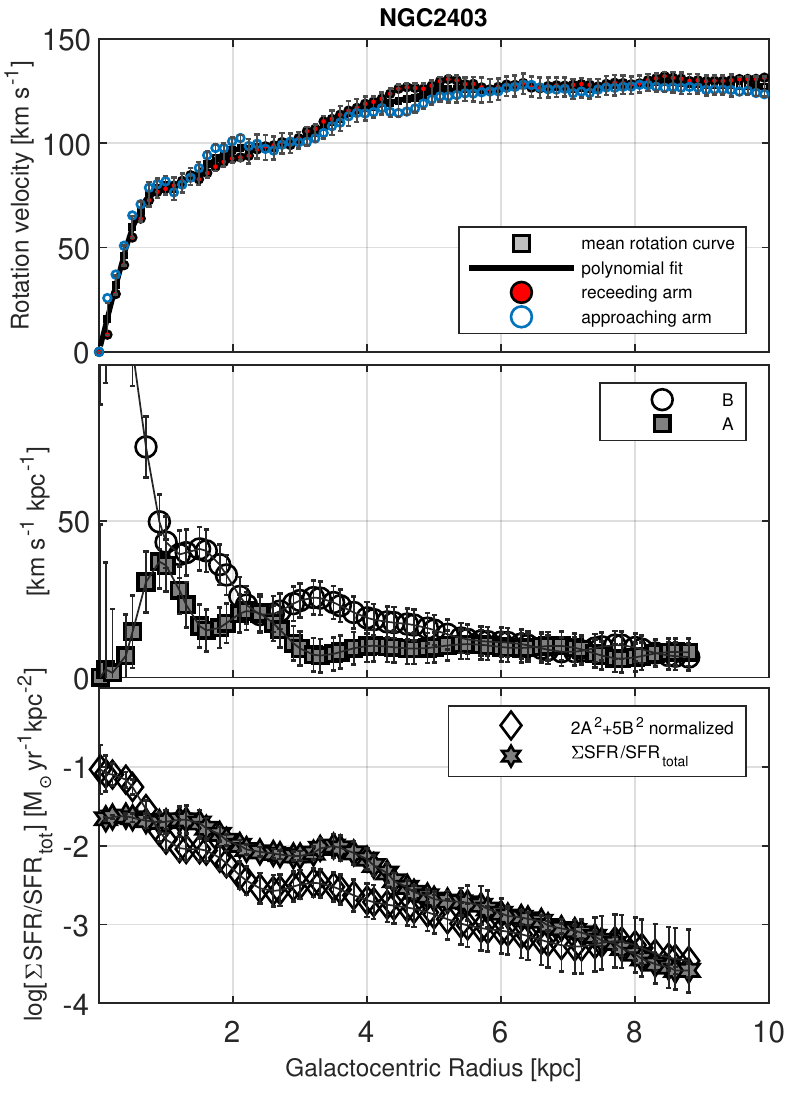}&\includegraphics[trim={0 0 0 0},clip,width=0.49\textwidth]{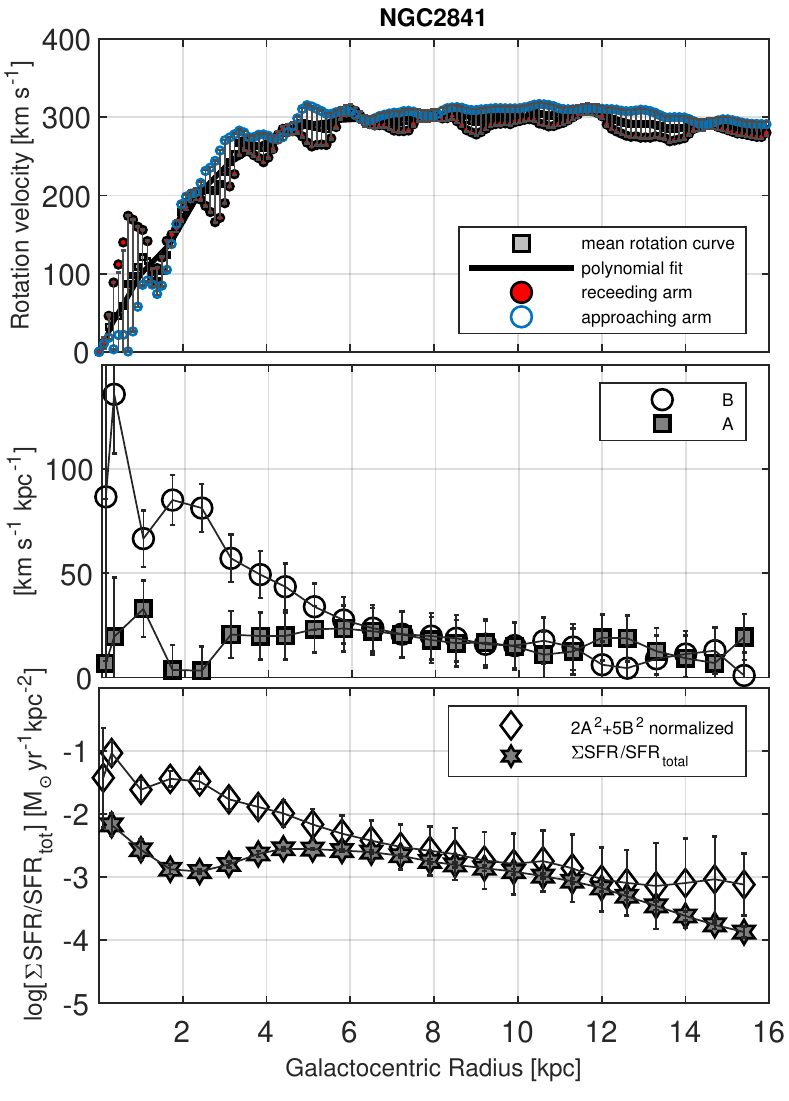}\\
    
	\end{tabular} 
	\caption{The upper panel of each figure shows the rotation curve in km~s$^{-1}$, the middle panel shows the 2 values of $A$ and $B$ in km~s$^{-1}$~kpc$^{-1}$ and the bottom panel  $log\frac{\Sigma_{SFR}}{SFR_{total}}$ with the value of $log [2 A^2+5 B^2]$ normalized to arbitrary units all as a function of galactocentric radius in kpc on the X axis}  
	\label{fig1}
	\end{figure*}

\begin{figure*}
	\begin{tabular}{ll}
    \includegraphics[trim={0 0 0 0},clip,width=0.49\textwidth]{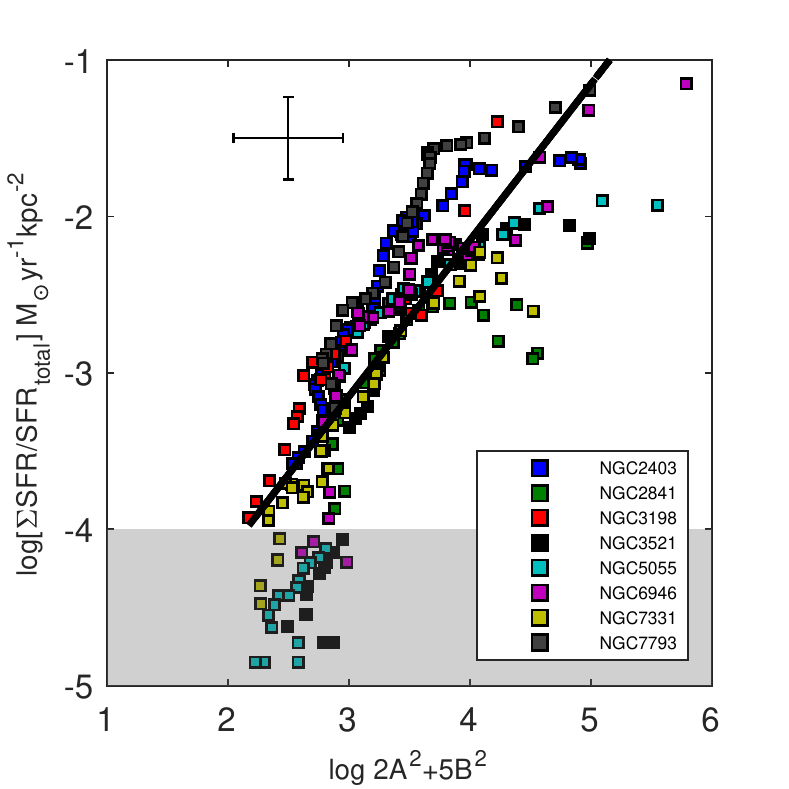}&\includegraphics[trim={0 0 0 0},clip,width=0.49\textwidth]{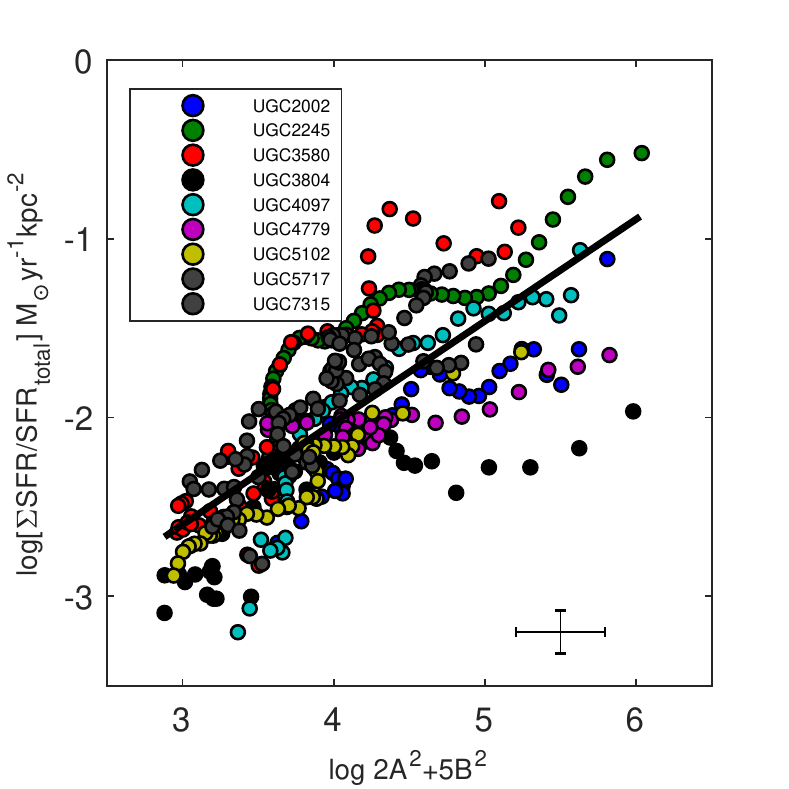}
    \end{tabular}
    \caption{$log\frac{\Sigma_{SFR}}{SFR_{total}}$ for each galaxy on the $y$ axis and the value $log2A^2 + 5B^2$ on the $x$ axis. The left panel panel is for the radio data and the right panel for the optical data. The black line is the best fit. The shaded area represents the low star formation regime.}
    \label{fig2}
	\end{figure*}

\section{Discussion}

\subsection{The relation of $\Sigma_{\mathrm{SFR}}$ to A and B}
\begin{figure*}
\includegraphics [ width=1\textwidth]{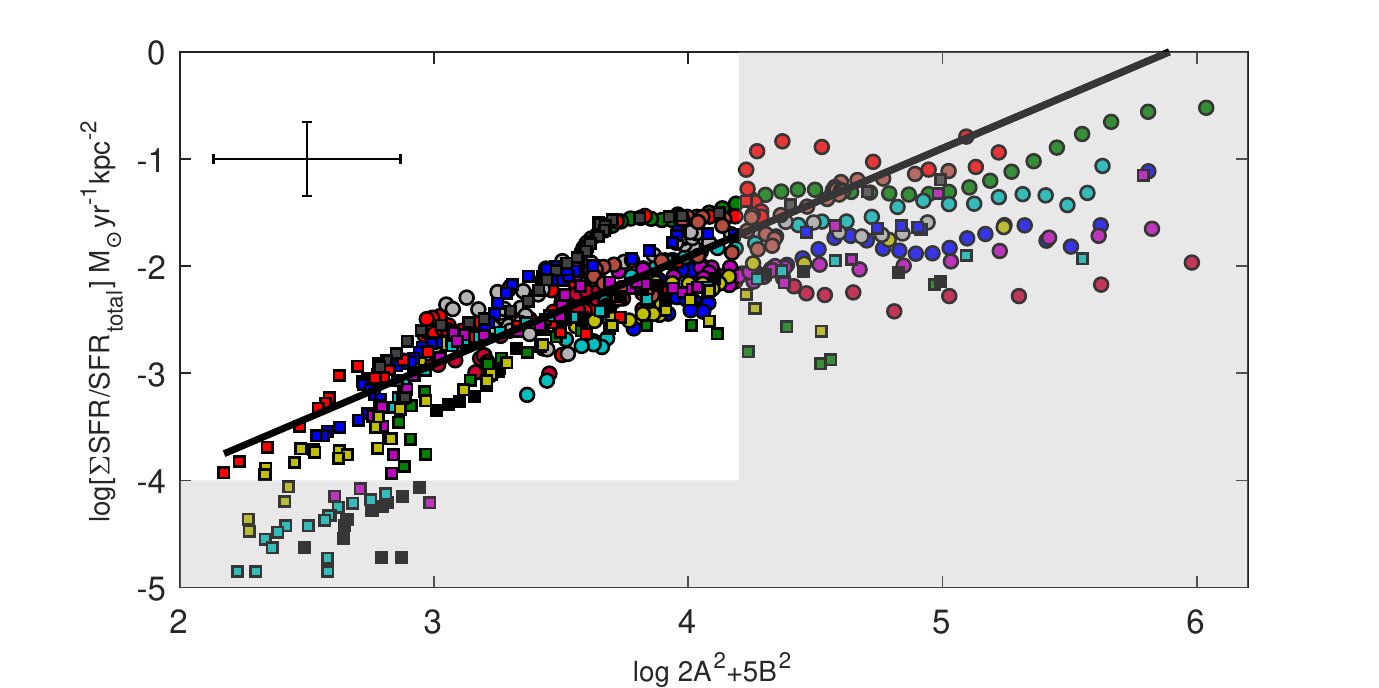}
\caption{$log\frac{\Sigma_{SFR}}{SFR_{total}}$ for all data points of all galaxies on the $y$ axis and the value $log2A^2 + 5B^2$ on the $x$ axis. The black line is the best fit to the data. The shaded area represents the data set excluded from the best fit. Colors and symbols are similar to Fig. \ref{fig2}.}
\label{fig3}
\end{figure*}

The trends we find in Figs.~\ref{fig1}, \ref{fig2} and \ref{fig3} are not expected if the star formation process was only due to the growth rate of density fluctuations in the disk. If this was the case, we should expect the differential rotation to act to regulate the process and suppress the star formation.
Areas of strong shear should create stability against collapse; however, this does not show in our data. 
Additionally, areas of strong vorticity quantified by $B$ are not expected to follow the trend of the star formation rates. However, the star formation rate increases with vorticity.
The trend found in this study seems to indicate that another process controls the star formation mechanism, which is enhanced with differential rotation rather than suppressed. 
This is possible if the clouds instead of being destroyed by shear, or supported by local vorticity, are on the contrary colliding in regions of strong shear.

We consider a frame of reference that is rotating at the same angular frequency $\Omega$ of the galaxy at any galactic radius $R_0$, with the \emph{x} axis pointing radially outward, the \emph{y} axis in the direction of rotation and the \emph{z} axis perpendicular to the disk. In nearly circular orbits, relative motions of fluid parcels in the \emph{y} coordinate are caused by the differential rotation, while motions in the \emph{x} and \emph{z} coordinates are caused by radial and vertical perturbations. These perturbations can be due either to collisionless encounters (scattering by other giant molecular clouds) or by the azimuthally varying gravitational potential  and they will result in epicyclic motions  with associated velocity components and amplitudes. \citet{spitzer1951} argued that the  stellar velocity dispersion is caused by such collisionless encounters of stars with massive molecular clouds. However, the cloud--cloud scattering is different from the star--cloud scattering in the sense that the epicyclic amplitude for the former is of the same range as the cloud tidal radius while for the latter it is much larger. Therefore, the interaction velocities between two different clouds will be dominated by the contribution from differential rotation (shear dominated encounters), rather than by dispersion (dispersion dominated encounters) \citep{Binney2008}.

The same approximation was used in a number of studies \citep[e.g.][]{Gammie1991,Tan2000}. They consider that the resulting velocity of collision between clouds is several times larger than the bulk random velocities due to scattering. These authors consider that the  encounters between clouds happen at impact parameters of $\sim$1--2 tidal radii and the velocity of these encounters is the  shear velocity due to differential rotation. The role of the velocity dispersion is to set the  clouds on epicycles, but it is not the velocity influencing the collision rates. The effect of velocity dispersion is to increase the impact parameter at which the clouds collide and therefore increases the frequency of collisions and not the velocity of encounters.

The tidal radius is defined as the radius at which the escape velocity from the cloud is equal to the shear velocity, and represents the volume where the gas is confined gravitationally to the mass contained within its radius rather than escaping due to differential rotation. Giant molecular clouds have been observed to have a radius that is on the order of their tidal radius \citep{stark1978}. \citet{Gammie1991} and \citet{Tan2000} use fiducial values in the range of $\sim$100 pc, and thus, $\sim$1--2 tidal radii is consistent with the scales discussed here.

If we think of the sheared disk as different annuli sliding on each other, then the relative velocity between these annuli is the shear velocity \.y (in the \emph{y} coordinate).
We neglect motions in the \emph{z} axis  because the size of the clouds is similar to the scale-height of a disc \citep{stark1978,solomon1987,Tan2000,guo2015} and the GMC occupy a thin vertical spread giving them a two rather than a three dimensional distribution.

Moreover, the vorticity $\omega$ measures the tendency of parcels of the fluid to rotate around a central point, hence, the differential rotation  will cause  gravitationally bound clouds to rotate around their centres of mass. The angular frequency of this local rotation  will be proportional to the second Oort parameter B at that point with $\omega=\vert2B\vert$. If gravitationally bound clouds inherit their angular momentum from the galaxy rotation, the radius below which they will rotate should in principle be similar to or smaller than their tidal radius. 

The positive gradients of the rotation curve will result in prograde rotations of the clouds while negative gradients will result in retrograde rotation of the clouds. In fact both of these trends have been reported in observations \citep{2Rosolowsky2003,Braine2019}. One critical point to note here is that the value of $B$ measures the tendency of the clouds to rotate, however it is not a direct measurement of their rotation. Numerous observational studies provide evidence that clouds rotate and there is some evidence that their rotation is due to the differential rotation of the galaxy. \citet{Kutner1977} observe two giant molecular cloud complexes in the Orion region and they measure a large scale velocity gradient of 135\,km~s$^{-1}$~kpc$^{-1}$ , hence they argue that the source of this angular momentum is the galactic rotation. \citet{2Rosolowsky2003} observe 45 GMC in Messier~33 with a resolution of 10~pc, and report velocity gradients in the range of 100\,km~s$^{-1}$~kpc$^{-1}$ which they interpret as an indication of rotation. Furthermore, they report that 40~per~cent of the clouds are counter rotating with respect to the galaxy. 
\citet{Braine2018} observe 566 clouds in Messier~33 and find that the majority of the clouds have rotational frequencies around 50~km~s$^{-1}$~kpc$^{-1}$.
\citet{Braine2019} observe 1058 clouds in Messier~51 on resolution scales of $\sim40$ pc and argue that the velocity gradients indeed reflect clouds rotation and their gradients follow galactic shear. Additionally the rotation of the clouds is prograde in rising parts of the rotation curve, which indicates that the rotation of the clouds is largely inherited from the rotation of the galaxy.
 
If clouds collide in areas of strong shear, then from a non-inertial frame of reference, i.e. centred on one of the clouds,  the energy of collision will be the sum of the energy of motion in the \emph{y} coordinate (with velocity proportional to $A$) and the energy of rotation (with angular frequency proportional to $B$). The total kinetic energy of a moving, rotating cloud in a differentially rotating disc will thus be 
 \begin{equation} 
 KE= \frac{1}{2}M_c \dot{y}^2+\frac{1}{2} I\omega^2 
\end{equation}
where $\omega=\vert2B\vert$ is the angular frequency of the local rotation, $\dot{y} = b (\Omega-\frac{dV}{dR})=2\times r_t \times A$ is the shear velocity at one tidal radius and $I=\frac{5}{2}M_c r_t^2$ is the moment of inertia of a rotating sphere, $M_c$ being the mass of the cloud and $r_t$ its tidal radius. The assumptions that the GMCs have a spherical geometry and a constant density distribution are a major simplification, but the change of the moment of inertia of the cloud geometry will not have a large impact on the result.
Replacing in equation (9) for the values $\dot{y}$, $I$ and $\Omega$ and dividing by $M_c$ we get the total energy of rotation per unit mass $KE$ :
\begin{equation} 
 KE= {r_t}^2(2{A}^2 +5{B}^2)
\end{equation}
If stars form through inelastic cloud collisions \citep{ricotti1997,tomokazu1965,guang2017} driven by shear velocities, then the rate at which they form should be proportional to the energy of such encounters, and we have demonstrated that this energy is proportional to the value of $2{A}^2 + 5{B}^2$.
As clouds collide at supersonic velocities, a shock is created at the collision interface compressing the gas further, creating dense clumps prone to further collapse and eventually star formation \citep{Anathpindika2009,anathpindika2010,anathpindika2009i}.
It is important to note the range over which this relation is linear. From Fig.~\ref{fig3}, it can be seen that it breaks down in the low star formation regime ($<-4$), shown in light grey shaded area. In fact as first shown by \cite{Bigiel2008}, a low-density sub-threshold regime exists, below which the dependence of the SFR density on gas density becomes uncorrelated.
It is also seen that this relation saturates at high kinetic energy and becomes non linear i.e. increasing the kinetic energy no longer increases the star formation linearly, which may be due to the fact that SFR cannot grow infinitely while in principle nothing limits the collision energy. 
Additionally, the empirical relation we find in equation \ref{eq8} can be a powerful tool; given a total star formation rate of a galaxy  and a rotation curve, (or a modeled gravitational potential) one can predict the star formation surface density distributions at different locations throughout the disk in rings of few hundreds of parsecs, without the need of gas data.

\begin{figure*}
	\begin{tabular}{ll}
    \includegraphics[trim={0 0 0 0},clip,width=0.49\textwidth]{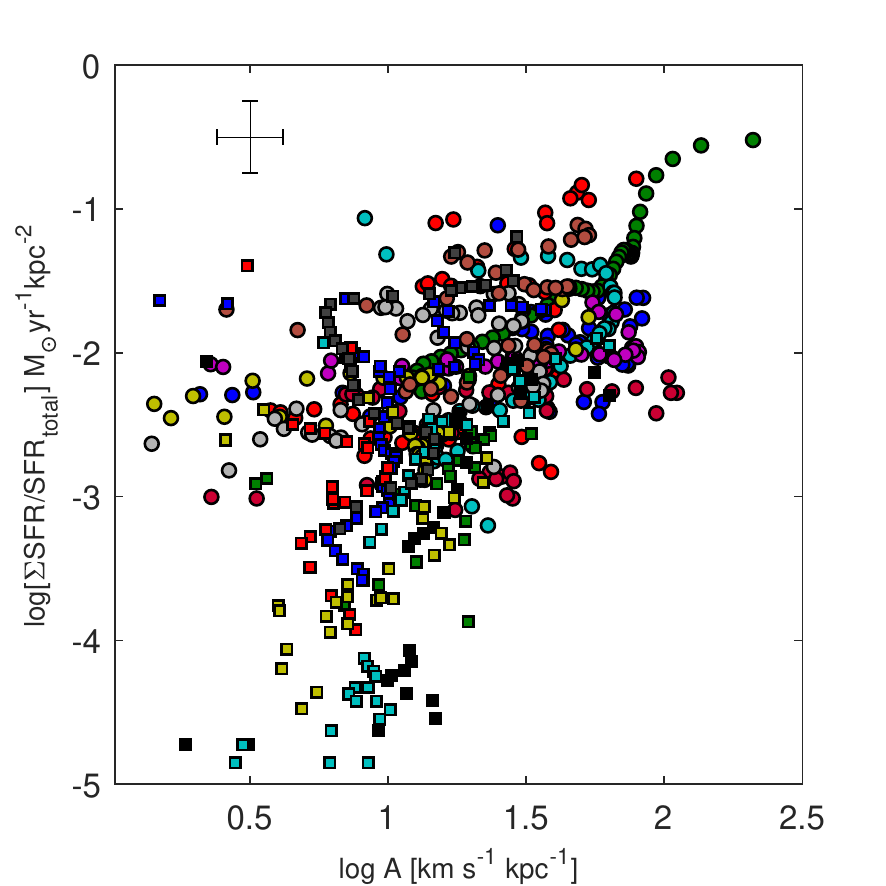}&\includegraphics[trim={0 0 0 0},clip,width=0.49\textwidth]{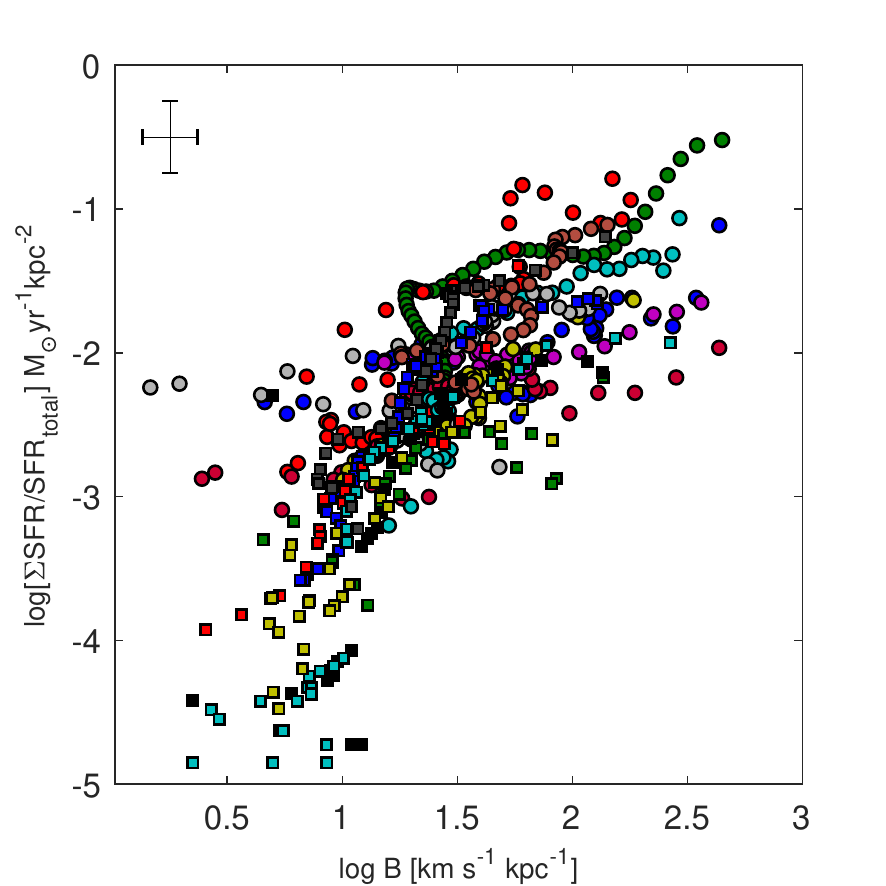}\\
    \includegraphics[trim={0 0 0 0},clip,width=0.49\textwidth]{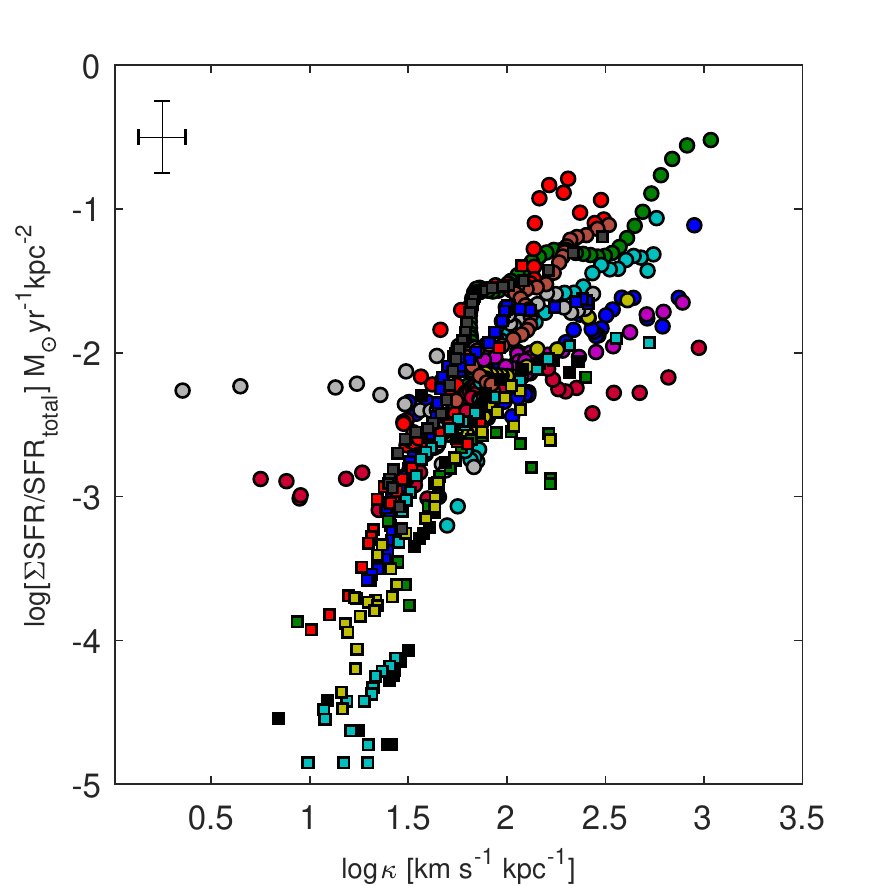}&\includegraphics[trim={0 0 0 0},clip,width=0.49\textwidth]{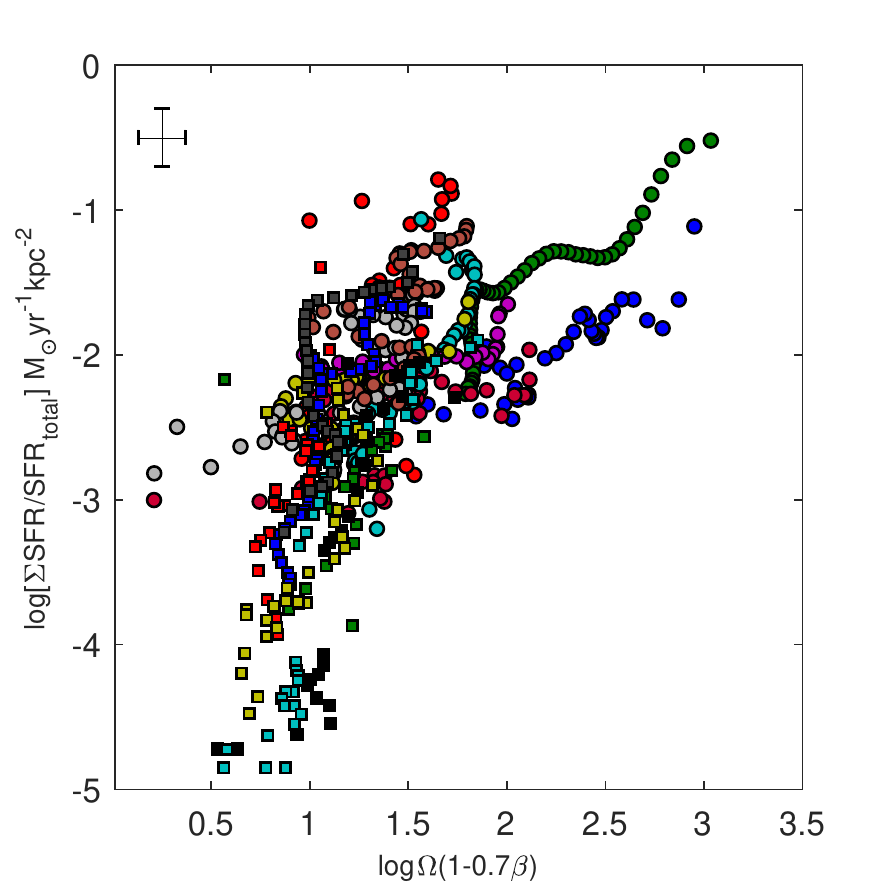}
    \end{tabular}
    \caption{$log\frac{\Sigma_{SFR}}{SFR_{total}}$ for each galaxy on the Y axis and the values of $log A$ (top left), $log B$ (top right), log $\kappa$  (bottom left) and log$\Omega(1-0.7\beta)$ bottom right. An increasing trend can be seen in which the star formation rate fraction distribution increases with the first Oort parameter $A$ and the value $\Omega(1-0.7\beta)$ both a measure of the local shear value. This trend is also visible with the second Oort parameter $B$ and the epicyclic frequency $\kappa$, both a measure of local vorticity.}  
    \label{fig4}
	\end{figure*} 

\subsection{Comparison with the Literature}

\subsubsection{Theory and Simulations}
The concept of cloud collisions has been previously proposed before \citep{Scoville1986,gilden1984,loren1976}. 
\citet{Tan2000}, motivated by the results of \citet{Gammie1991}, develops a theoretical framework for the cloud collision model. In this model, the GMC collisions cause the majority of star formation in the disk. He argues that these events are frequent in a differentially rotating disk ($\sim{5}$ per orbital time) and they drive the star formation rates observed in disk galaxies.
He predicts the star formation activity to be increased through cloud collisions in regions of strong shear and reduced  in regions of low shear. This is in agreement with recent simulations \citep{Tasker2009,dobbs2015}.
The exact model of Tan has the form of $\Sigma_{SFR}  \propto \Sigma_{gas}\Omega(1-0.7\beta)$
where $\beta=\frac{d_{\log V}}{d_{\log R}}$= $\frac{dV/V}{dR/R}$ = $\frac{R}{V} \frac{dV}{dR}$ and $\Omega=\frac{V}{R}$. Substituting we get:
\begin{equation} 
\Sigma_{SFR}\propto \Sigma_{gas}(\frac{V}{R}-0.7\frac{dV}{dR})\
\end{equation}
where the term $\frac{V}{R}- 0.7 \frac{dV}{dR}$ is a modified Oort parameter.
We plot the $\frac{\Sigma_{SFR}}{SFR_{total}}$ against the value $\frac{V}{R}-0.7\frac{dV}{dR}$ (Fig.~\ref{fig4} lower right panel) and find an increasing  trend, in accordance with the model of \citeauthor{Tan2000}.
The model of cloud collisions has been also proposed as an effective mechanism in which massive stars form \citep{Takahira2018,Fukui2014,habe1992}

Thus, if cloud collision is indeed an effective process to form stars, we should expect the efficiency of this process to depend  on the energy of such collisions and therefore their impact velocity.

Recent simulations have investigated the effect of varying the impact velocity of two colliding clouds with different masses. \citet{Fujimoto2014} finds that colliding clouds at velocities between 10~km~s$^{-1}$ and 40~km~s$^{-1}$ are successful 50 per cent of the times in producing stars, while velocities below and above these values are only successful at 5 per cent.
These limits correspond to values of $A$ between 33 and 132~km~s$^{-1}$~kpc$^{-1}$ if taken at ranges of 150 parsecs.
\citet{Takahira2018} perform simulations of cloud collisions, and vary the impact velocity of clouds between 5 and 30~km~s$^{-1}$. They find that faster collisions produce a larger number of cores and in shorter amount of time. Similar to \citet{Fujimoto2014}, they find that low speed values of 5km~$s^{-1}$ have less role in the star formation than speeds of 10-30 km~$s^{-1}$, these values correspond to a value of $A$ of 33~km~s$^{-1}$~kpc$^{-1}$ if considered at radial separation of 150 parsecs.
\citet{anathpindika2010} find that relative collision velocities less than 5~km~s$^{-1}$ between 2 clouds with masses 5000 and 2000 $M_{\odot}$ did not produce any cores. This corresponds to values of $A$ of 16~km~s$^{-1}$~kpc$^{-1}$.
On the observational side, an increasing number of studies report star formation regions at the interface of colliding clouds \citep{dewangan2018,enokiya2019,fukui2019,tsuge2019}. \citet{fujitaetal2019} observe two colliding clouds with respective velocities of 16~km~s$^{-1}$ and 25~km~s$^{-1}$ which makes their colliding velocity 8~km~s$^{-1}$. They claim that massive star formation at the collision interface is probably due to the collision between the clouds.

\citet{fukuietal2015} observe high mass star forming region in the LMC, and identify a shocked interface layer between two colliding clouds. The collision velocity is again 8~km~s$^{-1}$ and they argue that it is triggering the massive star formation.
\citet{Furakawa2009} and \cite{Ohama2018}  observe the super star cluster Westerlund 2 and they identify surrounding H{\sc ii} regions belonging to two distinct clouds. They argue that the triggering of this super cluster is the clouds' collision at a velocity in the range of 10~km~s$^{-1}$.
This presents some evidence that while high collision velocities are effective at triggering star formation, the process may not be as effective for lower impact velocities. In fact, if the main trigger of star formation is due to collision velocities due shearing motions, we should expect a decrease of their efficiency with lower values of shear and therefore $A$. 

Observations and simulations hint at a value of a minimum collision velocity affecting star formation of 10~km~s$^{-1}$, that corresponds to $A$ $\approx$ 30~km~s$^{-1}$~kpc$^{-1}$ at radial distances of 150~pc. 
These scales are consistent with our sampling and the resolution of our rotation curves.
In Fig.~\ref{fig4}, we plot the relation between the normalized star formation rate densities $\Sigma_{SFR}/SFR_{total}$ and $A$ (upper left panel), $B$ (upper right panel) and $\kappa$ (lower left panel) for 679 points from 17 galaxies. While the relation with $A$ shows an increasing trend, it is obvious that for lower values of $A$ the scatter increases, while for values larger than about $\log$1.5 the correlation is tighter. This is in accordance with simulations and may indicate a gradual shift of star formation regime, in the sense that the cloud-cloud collision model becomes more effective with increasing values of $A$ and it gradually decreases with decreasing values of $A$. At lower collision velocities some other factors will regulate star formation, thus increasing the scatter of the correlation.
The relations with vorticity $B$ and epicyclic frequency $\kappa$ show also an increasing trend with less scatter, indicating the importance of local rotational kinetic energy in the cloud collision process at these scales.

Simulations performed on the galactic scales reach varying conclusions.
\citet{Weidner2010} use simulations to argue that the presence of shear inhibits the formation of dense super clusters and predict the presence of these dense clusters only in dwarf galaxies and in interacting galaxies where the lack of rotation and therefore lack of shear provides stability against collapse. \citet{Hocuk2011} also present models of SFR in the vicinity of an AGN, finding that the star formation rate and efficiency are reduced when the shear level caused by the black hole increases in their models, although these simulations do not account for either magnetic fields or stellar feedback, both of which act to reduce the star formation activity.

\citet{Colling2018} perform a 3-D simulation of a 1~kpc cube in a differentially rotating disk including feedback from supernovae and H{\sc ii} regions. They deduce that galactic shear has an important influence on the star formation rate, in the sense that the higher the shear, the lower the SFR. It is worth to mention that this study considers a constant gradient to the rotation curve, and additionally they use in their simulations a 1~kpc cube, different from the flat disk approximation where the size of the GMCs is comparable to the disk scale height and where the clouds are considered to occupy a two- rather than a three-dimensional distribution.

Simulations performed by \citet{Fujimoto2014} show that the most massive clouds form in a high-interaction environment, namely in the dense bar region, to conclude that cloud collisions are an efficient way to build more massive clouds from smaller ones. They further test the star formation model of \citet{Tan2000} and find a slope of the KS law of $N\sim 2$ instead of 1.4 in accordance with the observations of \citet{HsiAn2014}.
Simulations performed of \citet{Tasker2009} and \citet{dobbs2015} find that the collision time scale is a small fraction (1/5) of the orbital time, in accordance with \cite{Tan2000}.

\subsubsection{Observations}
On the observational side \citet{Seigar2005}  finds a  strong and significant anti-correlation between the mean shear rate as calculated on the scale of the whole galaxy and the mean specific star formation rate (SSFR) in 33 galaxies, however he finds a weak relation between the shear rate and the star formation rate surface density. Similarly, \citet{Colombo2018} find a similar trend in which the $\Sigma_{sfr}$ decreases with increasing shear. It is important to note that both of these studies are performed on disk averaged quantities (the scale of the whole galaxy or kpc scales) and are not resolved to regions within the disk for both the star formation rates and for the rotation curves.
The shear rate definition $1-\beta \equiv \frac{A}{\Omega}$ that these authors  use is only valid  when assuming a constant gradient to the rotation curve (this definition of shear is a dimensionless parameter where the Oort parameter is normalized to the angular frequency $\Omega$) and it does not measure shear changes as a function of the galactic radius. While this reasoning may work on large scales, it fails when concerned with smaller scales; in fact the gradient of a rotation curve changes throughout the galactic disk and these changes affect significantly the shear velocities and vorticity.

\citet{Meidt2013} investigate the dynamics of gas in the inner 9 kpc of M51 and find that areas with low shear (as traced by the first Oort parameter $A$) are associated with low star formation efficiency while other areas with high shear are associated with high star formation efficiency. They hence argue that shear cannot be the agent preventing star formation in these locations and another source of stabilization is required.
\citet{Suwannajak2014}  use data from \citet{leroy2013} to test several star formation laws in the disks of 16 nearby galaxies  on scales of few hundreds of parsecs.
They find an increase of star formation efficiency per orbit in regions of strong galactic shear, consistent with the cloud-cloud collision model of \citet{Tan2000}. Our results echo the results of these two authors.\\*

In the context of the Milky Way, by analyzing the southern spiral arm on scales of 0.5 kpc, \citet{Luna2005} find that massive star formation is enhanced in areas of solid body rotation and reduced in regions of strong shear to conclude that shear is a limiting factor for star formation activity.
On smaller scales, \citet{Dibetal2012} investigate the effect of shear in the Milky Way on the scale of few parsecs (730 clouds with sizes from 0.2 to 35 parsecs), using the stability parameter derived by \citet{Hunter1998} and deriving the first Oort parameter from  azimuthally-averaged published rotation curve. They find no effect of shear on the star formation efficiency (SFE), i.e. the  star  formation rate surface density per gas surface density $\frac{\Sigma_{sfr}}{\Sigma_{gas}}$, and hence it is unlikely to be a dominant mechanism in determining the SFE of clouds. They argue that stellar feedback and magnetic fields may be more important stabilizing parameters than shear.
\citet{thilliez2014} use the same approach as \citet{Dibetal2012} to analyze giant molecular clouds  in the Large Magellanic cloud, and find no effect of galactic shear on the star formation efficiency. They conclude that once the GMC are formed, their star formation efficiency does not depend on the shear parameter or the galactic tidal effects.

It is important to note that the maximum value for $A$ that is considered by \citet{Dibetal2012} is $\sim$ 35~km~s$^{-1}$~kpc$^{-1}$  (that is the maximum value given by the rotation curve used at galactic radius of $\sim$ 3 kpc ). Our rotation curves show values much larger than this. Our results echo theirs as long as the values of $A$ are small, however this is not true for large values of $A$ where shear velocities are extreme. Furthermore, shear velocity depends on the radial ranges over which they are measured (the variable $b$ in equation \ref{eq2}). \citet{Dibetal2012} sample scales of a few tens of parsecs and at these distances shear velocity is insignificant and cannot account for encounter velocities that are sufficient to trigger star formation (on radial separations of 10 parsecs the shear velocity will be in the range of few~km~s$^{-1}$ which is much less than the velocity dispersion). In the current paper the scales range a few hundred of parsecs which causes the values of shear velocities to increase significantly. Therefore the scale at which the process is investigated is of major importance and studies done at different scales may lead to different results. In fact, quantities such as the gas density, the velocity dispersion and the star formation rate surface densities measured in the galaxy disk are not constant at any scale in the ISM and strongly depend on the distances over which they are measured \citep{elmegreenscalo2004,mckeeostriker2007,calzetti2012}. The ISM is clumpy and turbulent, and regions of star formation on small scales have much larger densities and different kinematics than the averaged densities and kinematics on large scales. \citet{calzetti2012} argue that both the slope and the scatter of the KS relation depend on the size of the sub-galactic region sampled. The well-known Toomre criterion is built on the assumption that the variables defining it are consistent and well defined on all scales, however this is not the case, and this led researchers to build a modified stability criterion taking into account the different scales \citep{romeo2010}.

\section{Conclusions}

We have investigated the effect of differential rotation on star formation rate surface density in the disks of 17 low redshift spiral galaxies, using 679 measurements on scales of few hundred parsecs.
We built rotation curves and extracted the first and the second Oort parameters $A$ and $B$, respectively a measure of shear and vorticity in a differentially rotating disk. We measured the star formation rate surface densities at matching scales, in azimuthally averaged rings again sampling scales of a few hundred parsecs.
We used two sets of data with different methods to derive our variables:
For the first set of data, the optical galaxies, the rotation curves were extracted using long slit spectroscopy and the star formation from narrow band H$\alpha$ photometry \citep{james2004}. 
The second set was taken from the literature  with rotation curves extracted using a tilted rings model from H{\sc i} Integral field spectroscopy from the THINGS galaxy survey \citep{walter2008} and star formation extracted using a combined data of FUV and NIR from \cite{Leroy2008}.

Our main results can be summarized as follows :
\begin{itemize}
\item We find an empirical relation linking the total star formation rate surface density distribution  in annular rings sampled on scales of few hundreds parsecs, and the Oort parameters $A$ and $B$  such that $\log\frac{\Sigma_{SFR}}{SFR_{total}} \propto log [2 A^2+ 5 B^2]$. 
\item We argue that the  star formation rates at these scales are driven by cloud-cloud collisions.  The term  $[2 A^2+ 5 B^2]$ is proportional to the total mechanical energy of collisions, where clouds collide at shear velocities proportional to $A$. Rotation takes place at angular frequencies proportional to $B$, i.e. driven by the differential rotation.
\item We find that for low values of $A$ the cloud-cloud collisions are not an effective trigger for star formation since the shear velocities are too low on scales \textless 100  parsecs.
\item This empirical relation can predict how the total star formation rate of the galaxy is distributed through the disk  knowing only a rotation curve without the need of gas data. It may be also used the other way around: knowing the star formation rate density distribution throughout the disk, one can predict a rotation curve. This relation can be taken in account in numerical simulations investigating the evolution of galaxies, or it can be used for distant galaxies for which the rotation curves are measured but the resolved star formation rates are not resolved.
\end{itemize}

Future work should include a much larger sample to investigate trends related to galaxy classification. Azimuthally averaged star formation rates smooth out the effect of density waves patterns and a rotation curve is only a mean value of the rotation velocity across the disk and does not describe spatially resolved velocities. One approach would be to investigate the effect of rotation in spatially resolved regions rather than azimuthally averaged quantities both for the star formation and for the rotation curves.
Our simplified  interpretation can be tested through dynamical models, and since the shape of the rotation curves is of crucial importance, simulations with varying shapes and different gradients should be tested. Furthermore, these models can show whether and on what scales dense clouds rotate within differentially rotating disks.

\section*{Acknowledgments}

We thank Fabien Walter and Erwin de Block for providing the THINGS rotation curves in a computer readable form. We also thank Charles Lada for useful discussions. We thank Sue Percival for obtaining and reducing the optical spectroscopy, Christina Schoettler for proof reading and useful comments and Adrian Jannetta for double checking some of our mathematical statistical analysis. IC's research is supported by the Smithsonian Astrophysical Observatory Telescope Data Center, the Russian Science Foundation grant 19-12-00281 and the Program of development at M.V. Lomonosov Moscow State University through the Leading Scientific School ``Physics of stars, relativistic objects and galaxies''.

The Isaac Newton Telescope is operated on the island of La Palma by the Isaac Newton Group in the Spanish Observatorio del Roque de los Muchachos of the Instituto de Astrof\'isica de Canarias.
This research has made use of the NASA/IPAC Extragalactic Database (NED) which is operated by the Jet Propulsion Laborato\
ry, California Institute of Technology, under contract with the National Aeronautics and Space Administration.
This work made use of \say{THINGS},"The H{\sc i} Nearby Galaxy Survey" \citep{walter2008}.

\bibliographystyle{mnras}
\bibliography{references}

\appendix
\section{}
\label{appendixA}

The upper panel of each figure represents the rotation curve in ~km~s$^{-1}$, the blue dots and the red dots represent the receding and the approaching arm respectively, the squares are the mean values and the black line is a polynomial fit of varying degrees using a least square fit method. The second panel represents the two Oort parameters $A$ and $B$ in ~km~s$^{-1}$~kpc$^{-1}$. On the third panel we plot the fraction of the star formation rate surface density (the star formation rate in annular rings of few hundred of parsecs normalized to the total star formation rate of the galaxy) in grey filled stars $\log\frac{\Sigma_{SFR}}{SFR_{total}} $with the value $log [2 A^2+ 5 B^2]$ in empty diamonds. The $x$ axis represents the galactic radius in kpc.

\begin{figure*}
	\begin{tabular}{ll}
    \includegraphics[trim={0 0 0 0},clip,width=0.49\textwidth]{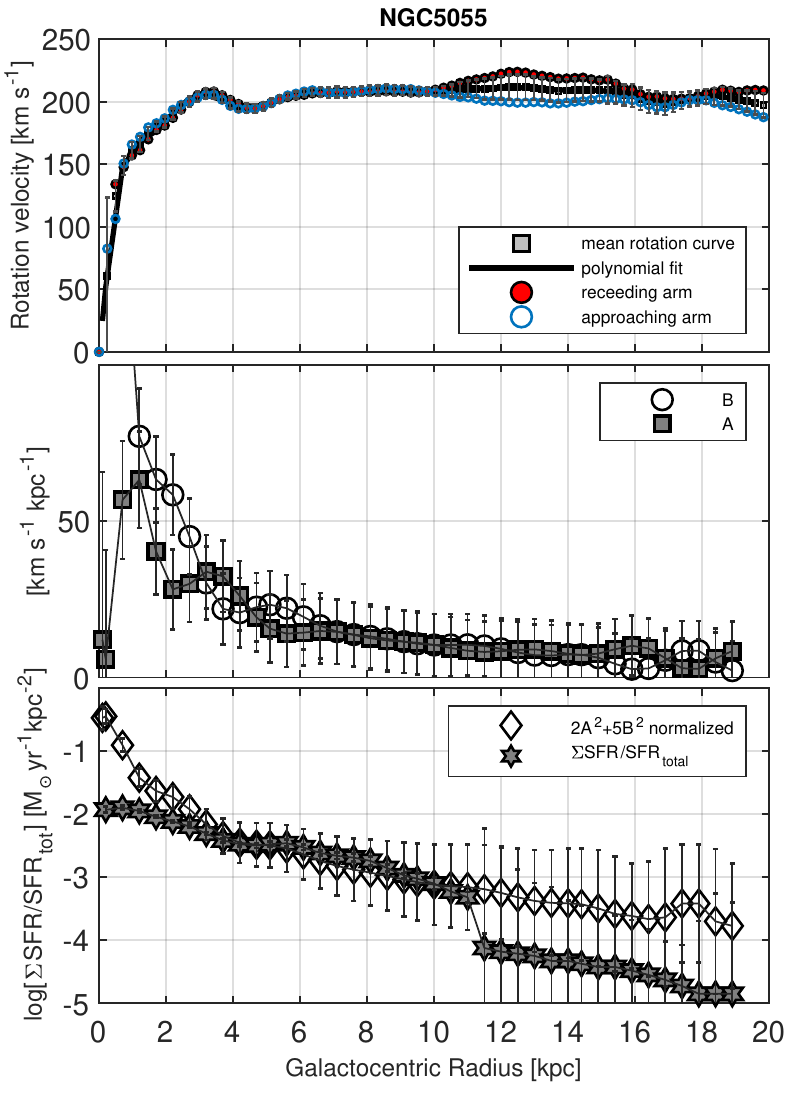}&\includegraphics[trim={0 0 0 0},clip,width=0.49\textwidth]{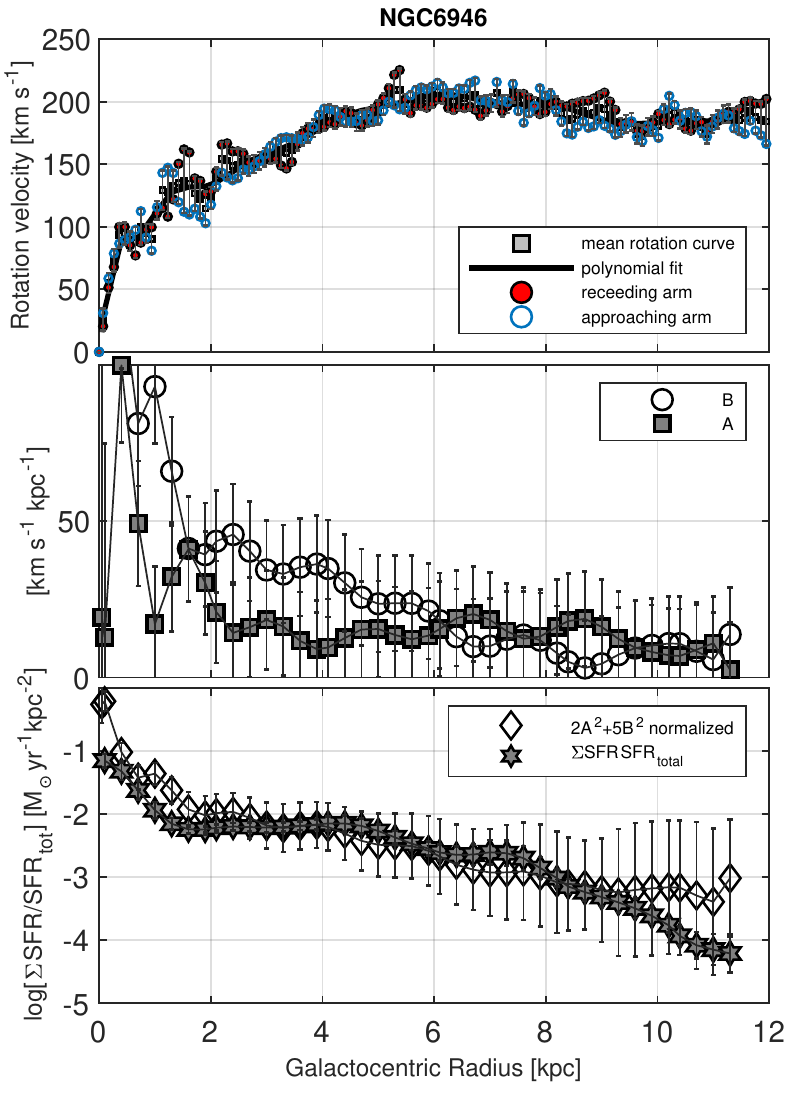}\\
    \includegraphics[trim={0 0 0 0},clip,width=0.49\textwidth]{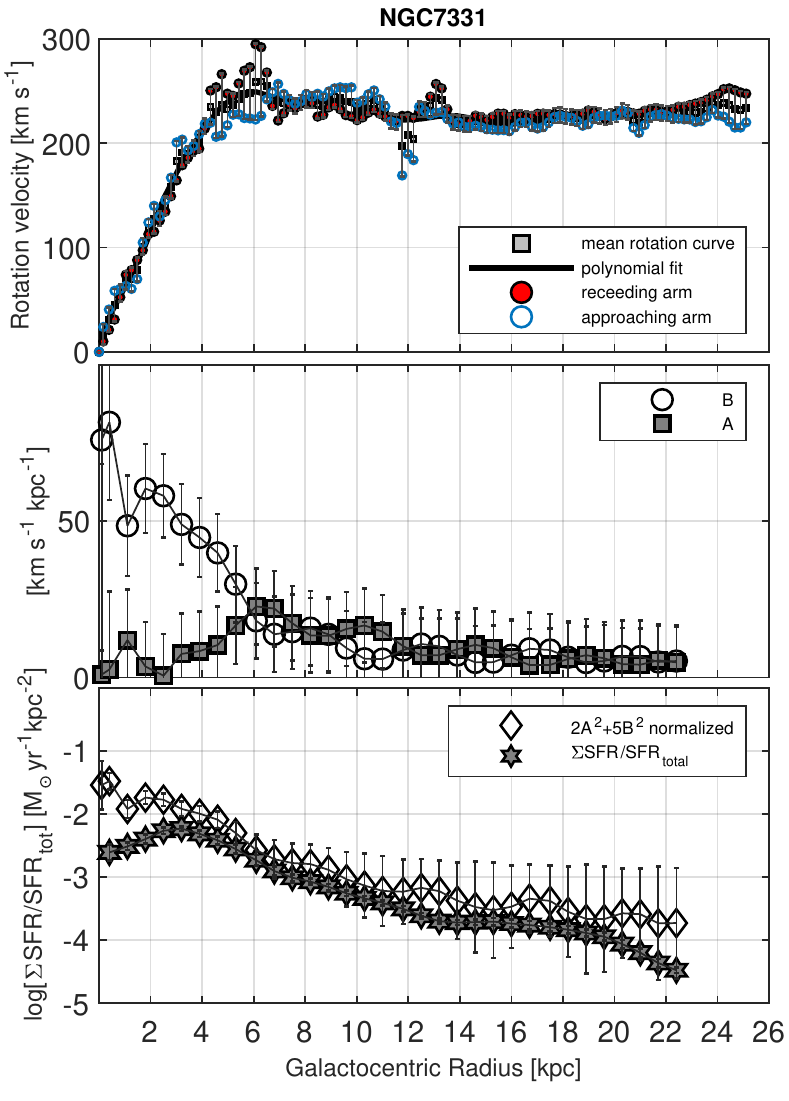}&\includegraphics[trim={0 0 0 0},clip,width=0.49\textwidth]{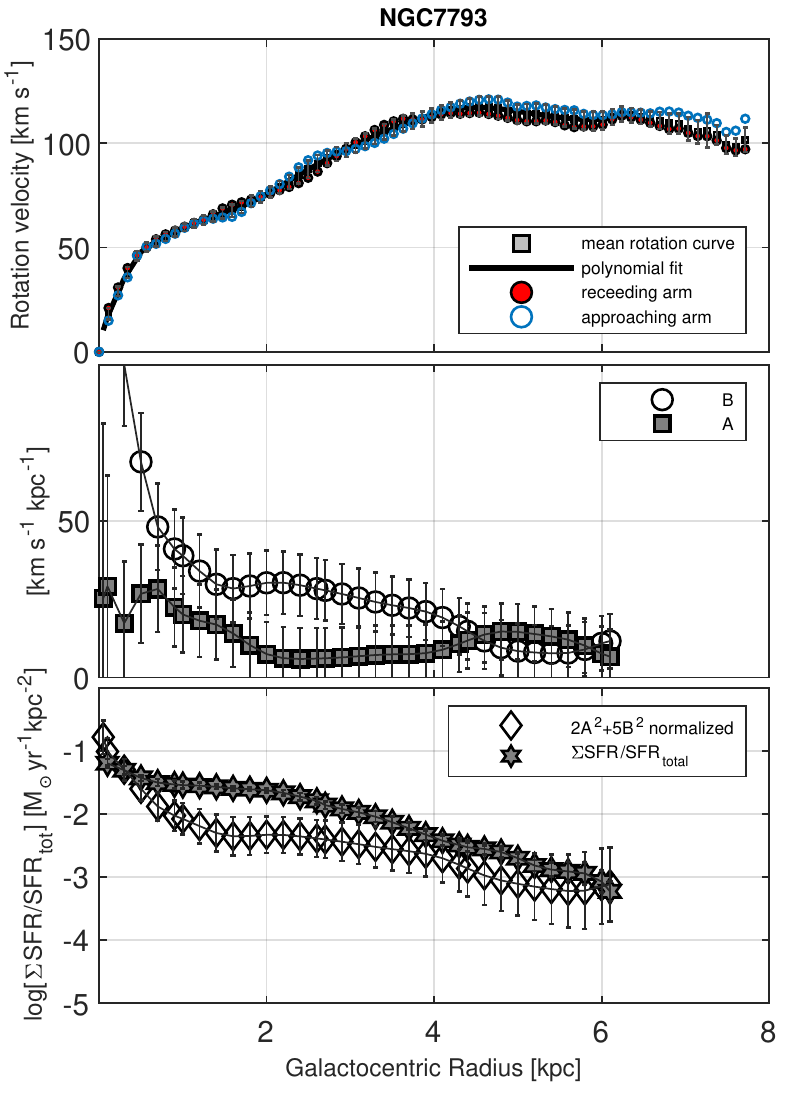}
	\end{tabular}
\end{figure*}

\begin{figure*}
	\begin{tabular}{ll}
    \includegraphics[trim={0 0 0 0},clip,width=0.49\textwidth]{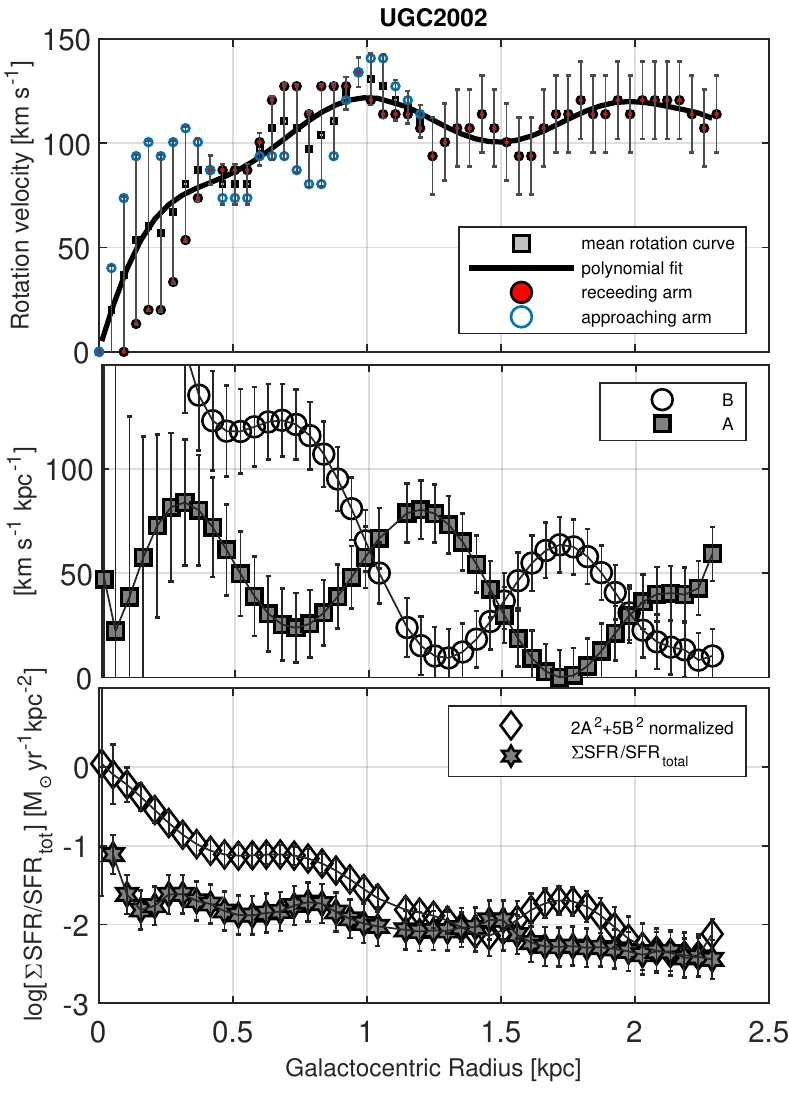}&\includegraphics[trim={0 0 0 0},clip,width=0.49\textwidth]{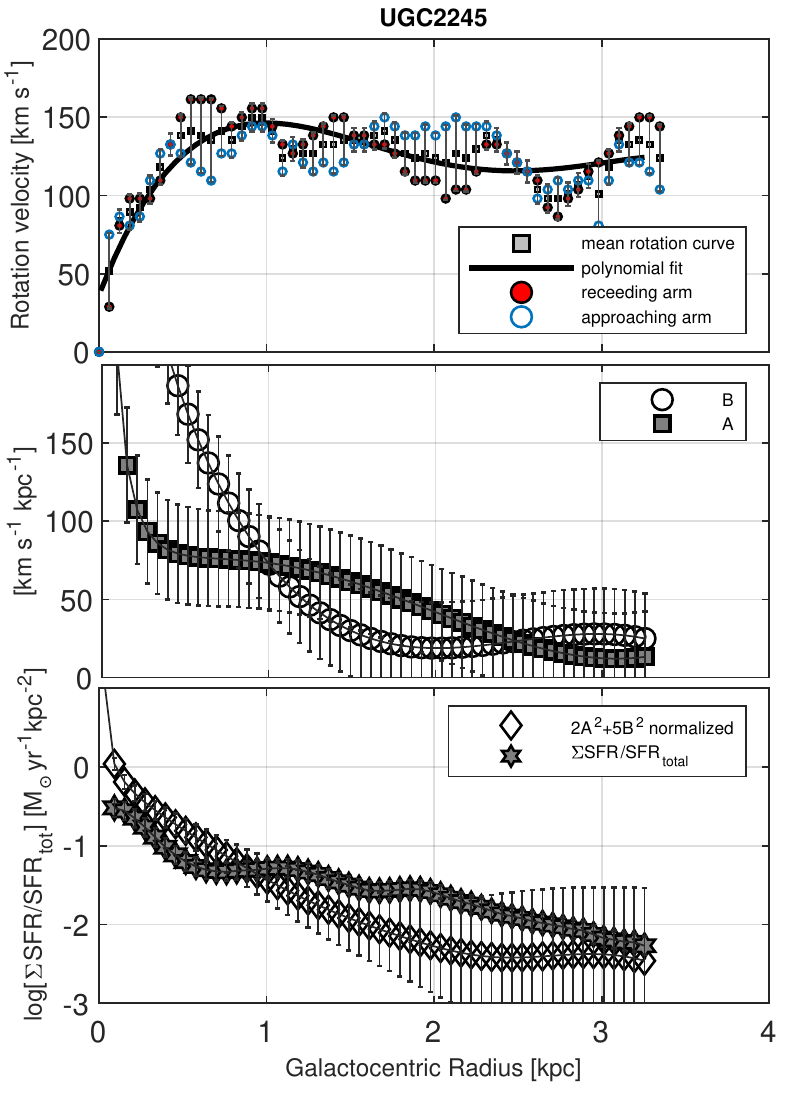}\\
    \includegraphics[trim={0 0 0 0},clip,width=0.49\textwidth]{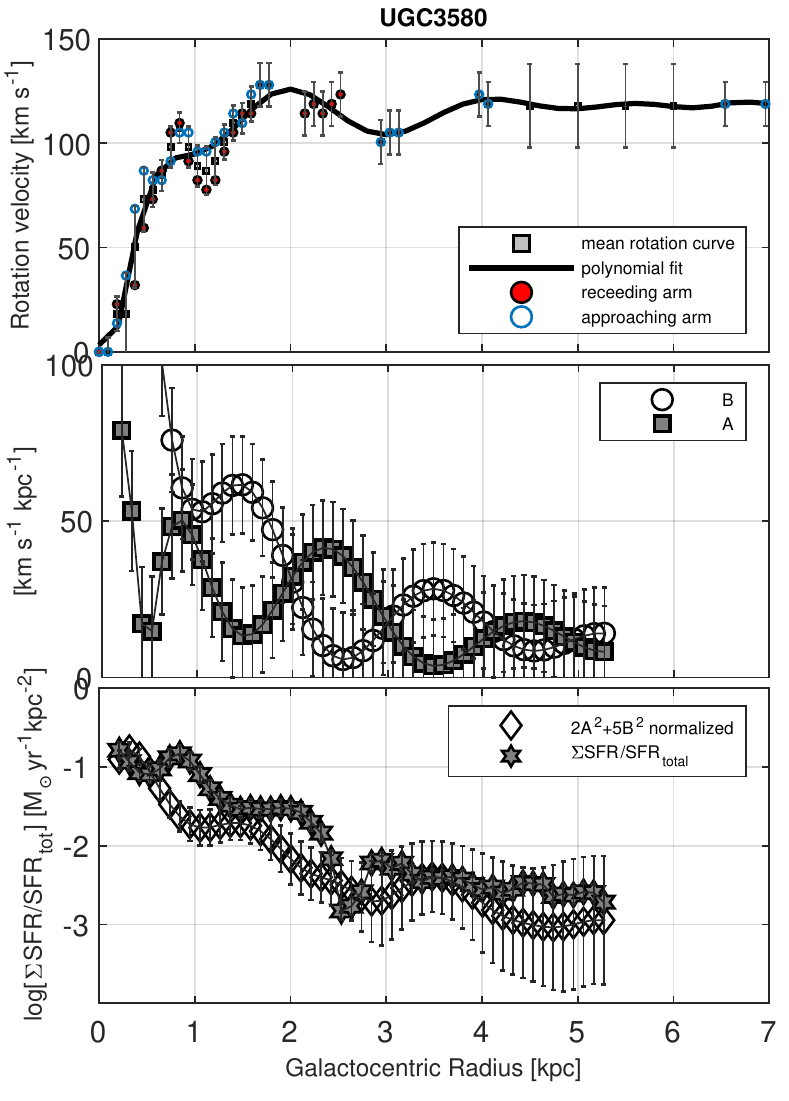}&\includegraphics[trim={0 0 0 0},clip,width=0.49\textwidth]{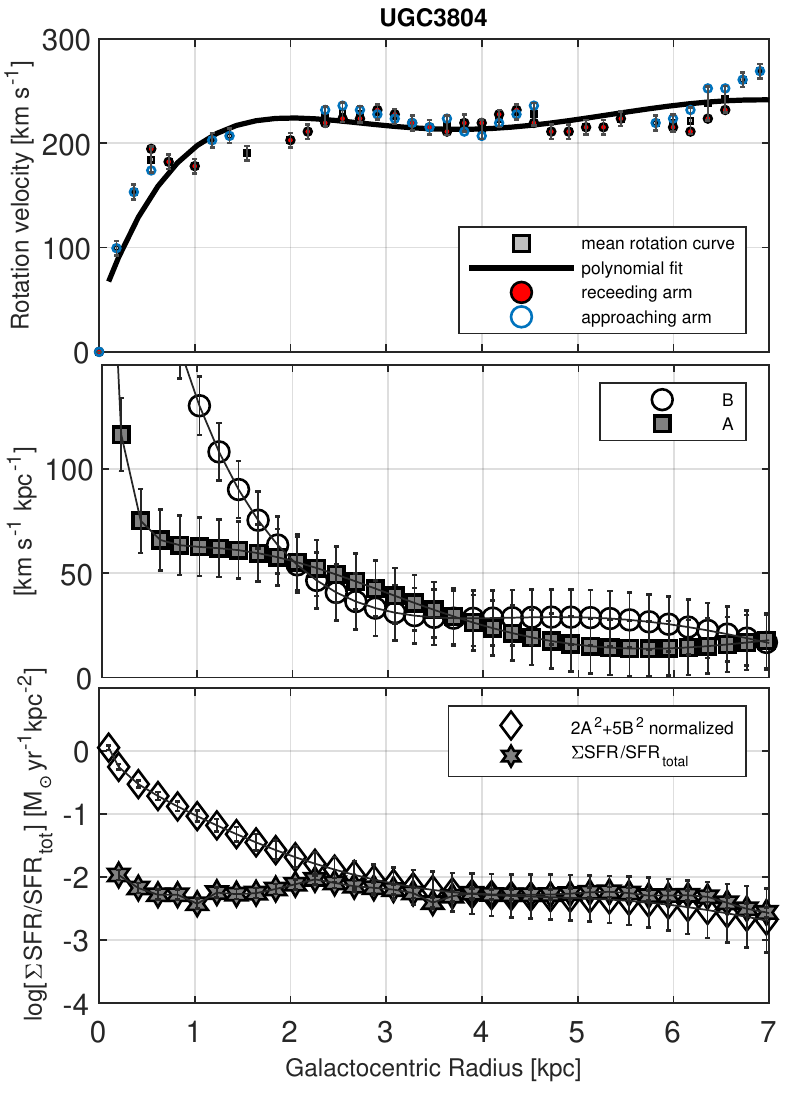}
    	
	\end{tabular}
	\end{figure*}

\begin{figure*}
	\begin{tabular}{ll}
    \includegraphics[trim={0 0 0 0},clip,width=0.49\textwidth]{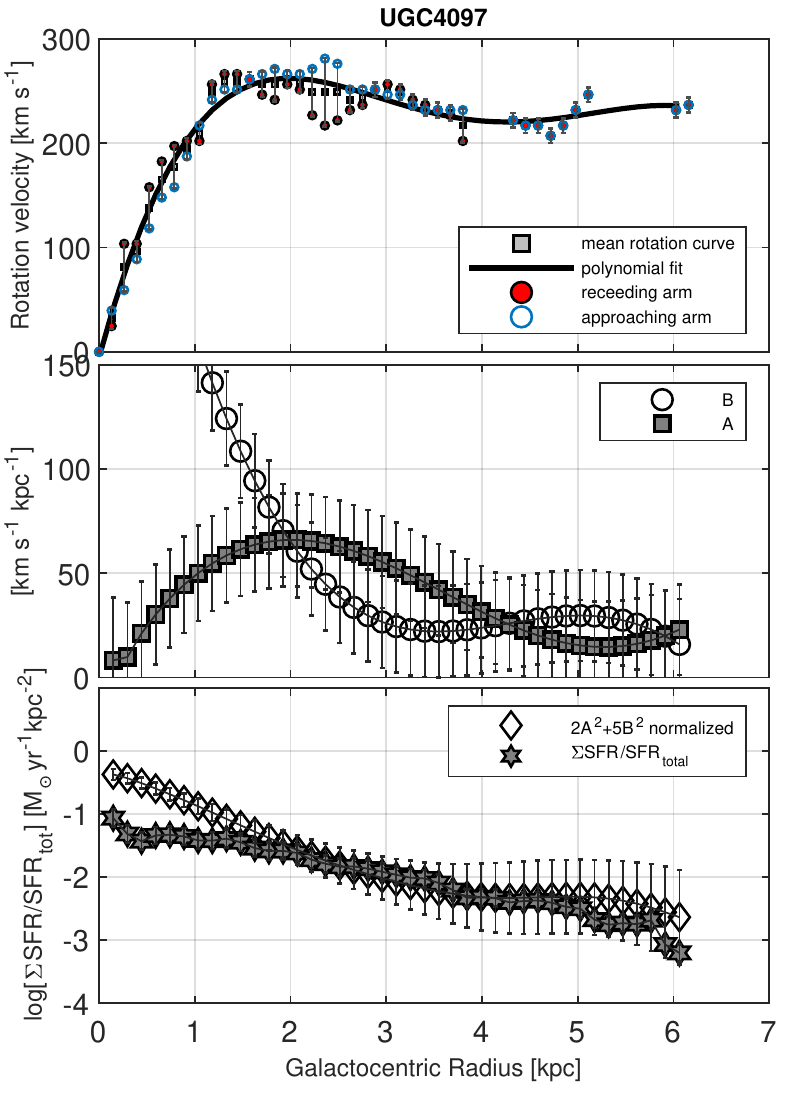}&\includegraphics[trim={0 0 0 0},clip,width=0.49\textwidth]{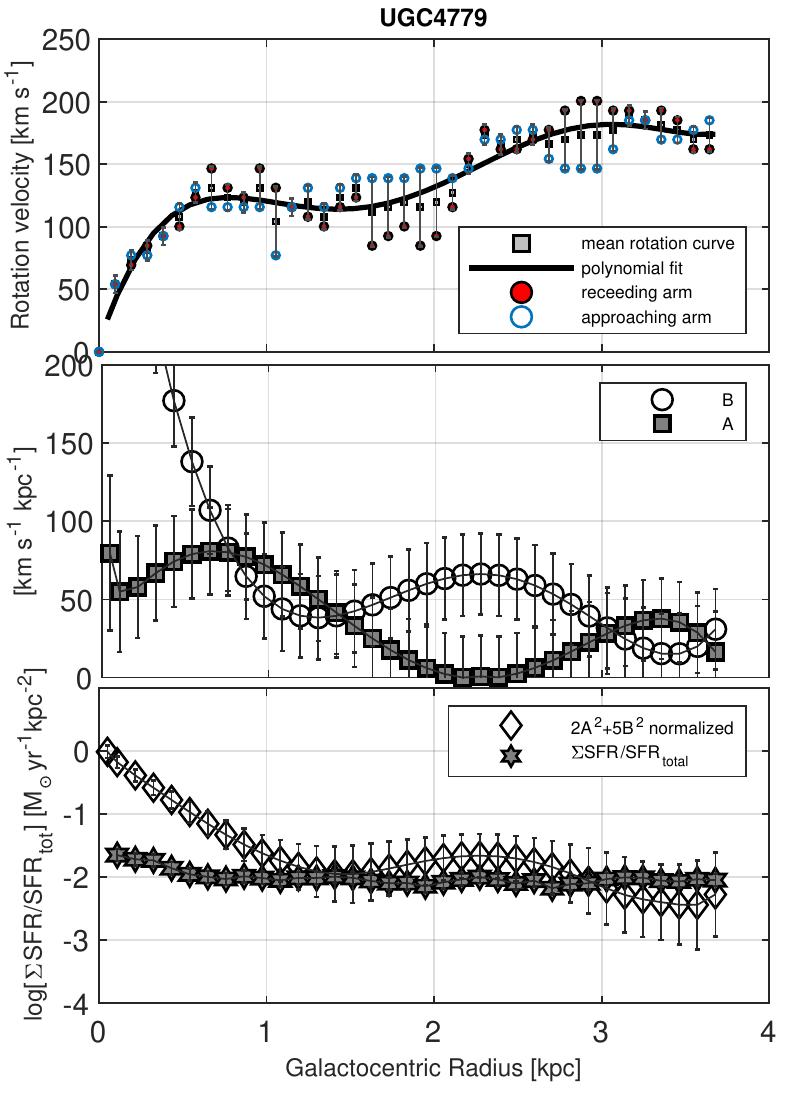}\\
    \includegraphics[trim={0 0 0 0},clip,width=0.49\textwidth]{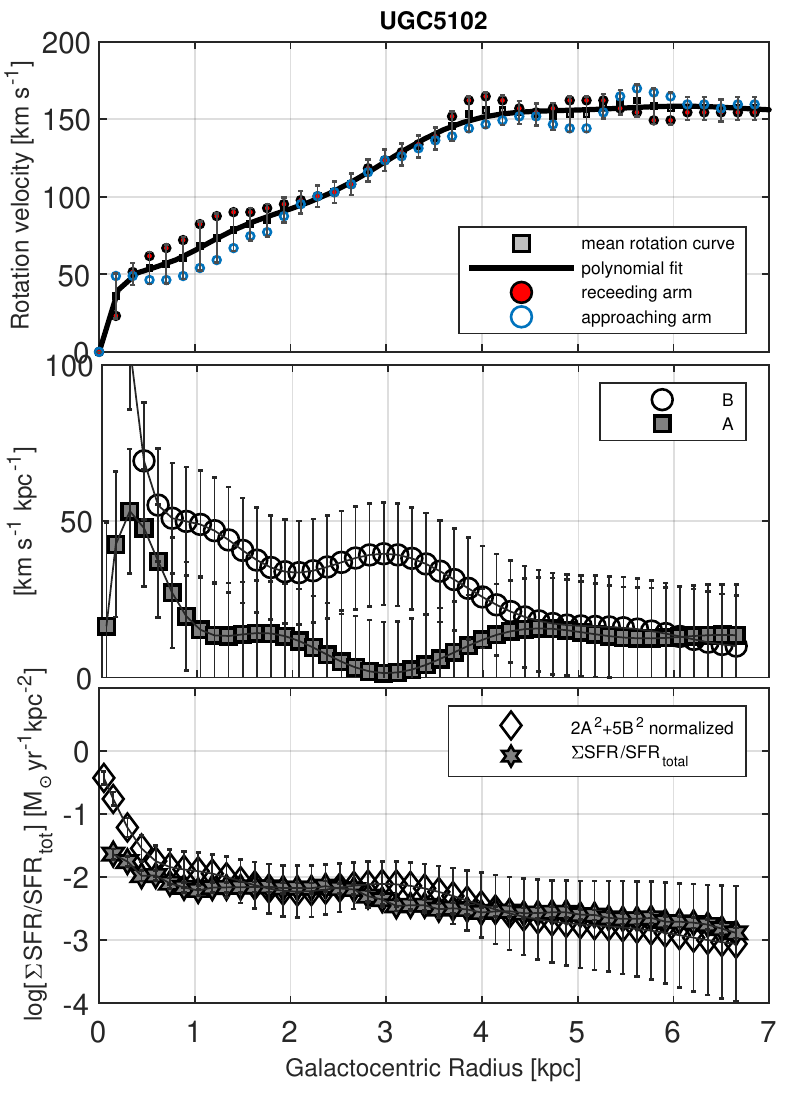}&\includegraphics[trim={0 0 0 0},clip,width=0.49\textwidth]{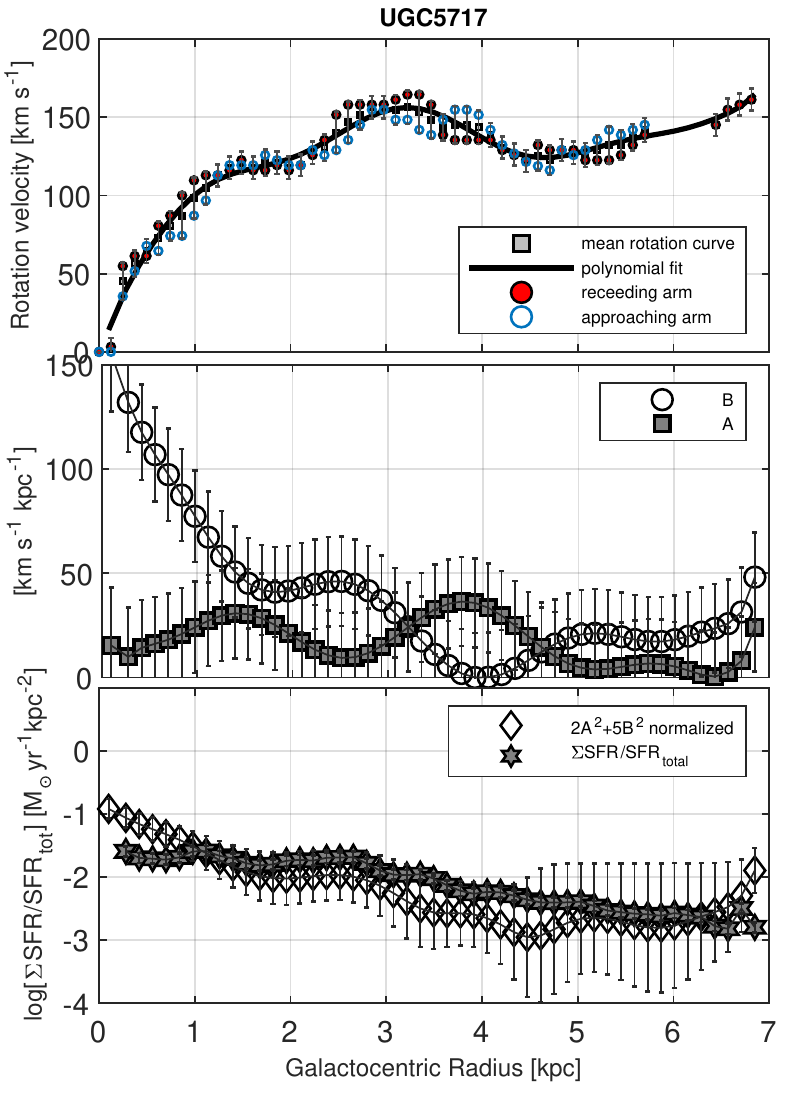}
	\end{tabular}
	\end{figure*}
	
\begin{figure*}
	\begin{tabular}{c c}
\includegraphics[trim={0 0 0 0},clip,width=0.49\textwidth]{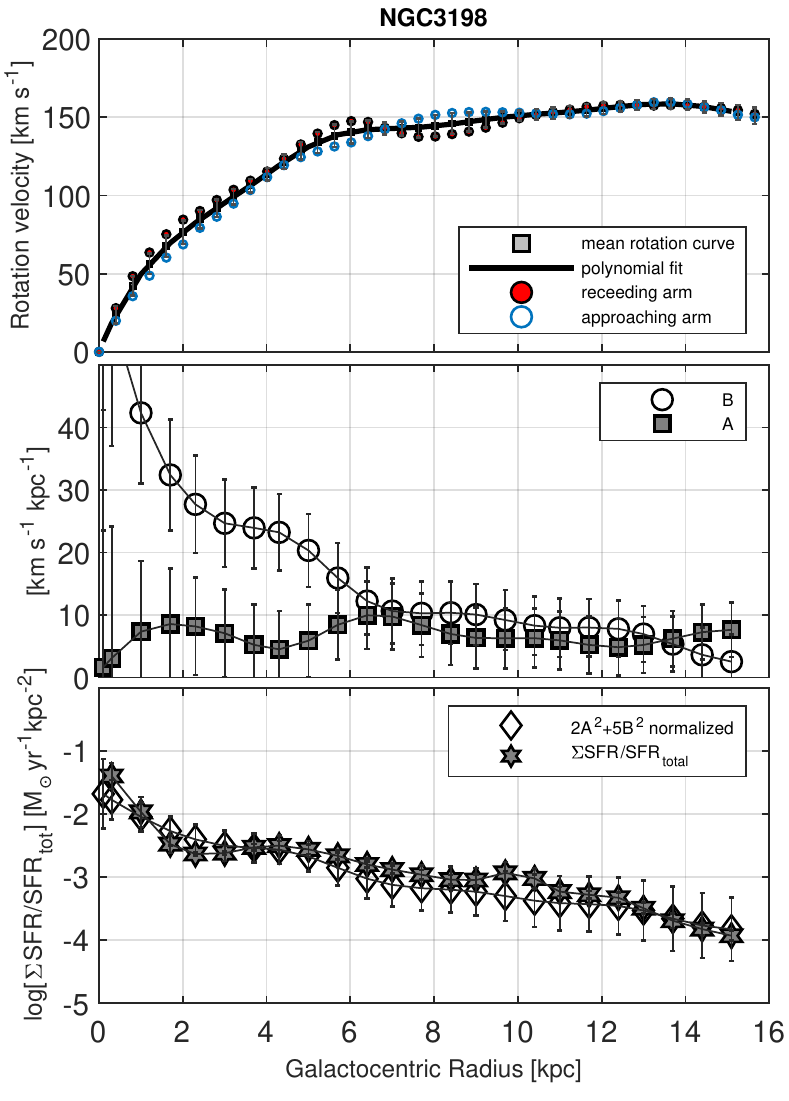}&\includegraphics[trim={0 0 0 0},clip,width=0.49\textwidth]{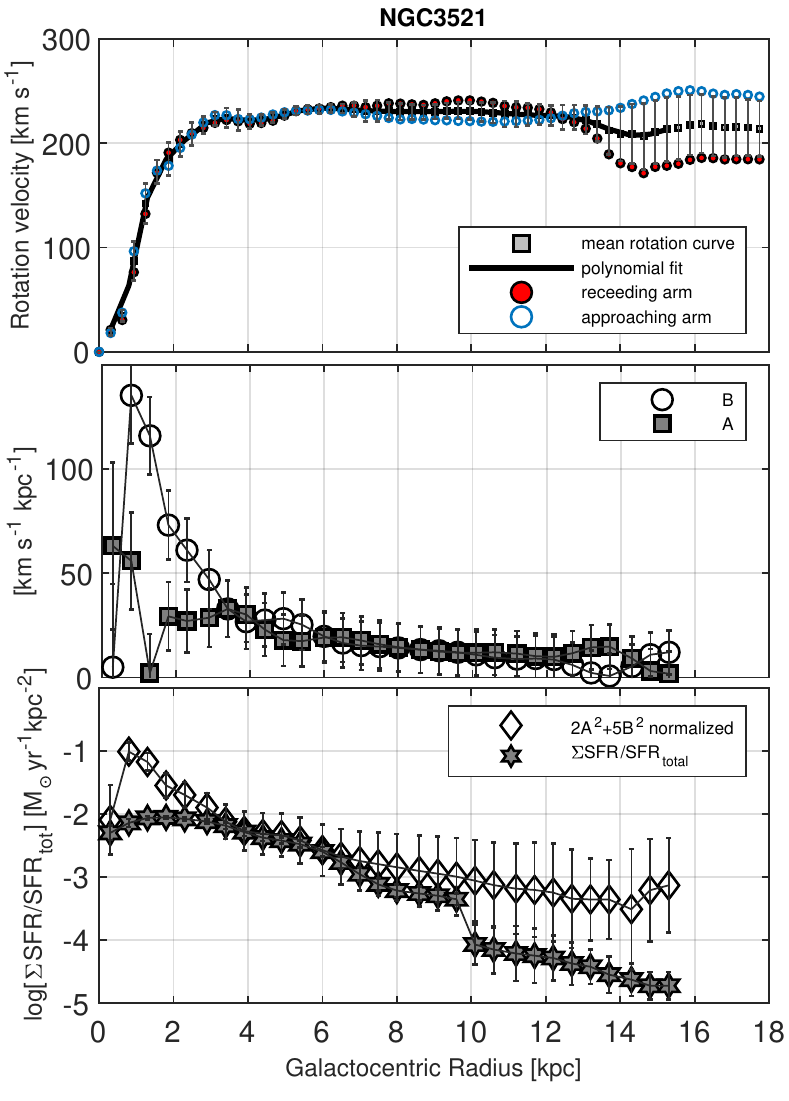}\\
 \includegraphics[trim={0 0 0 0},clip,width=0.49\textwidth]{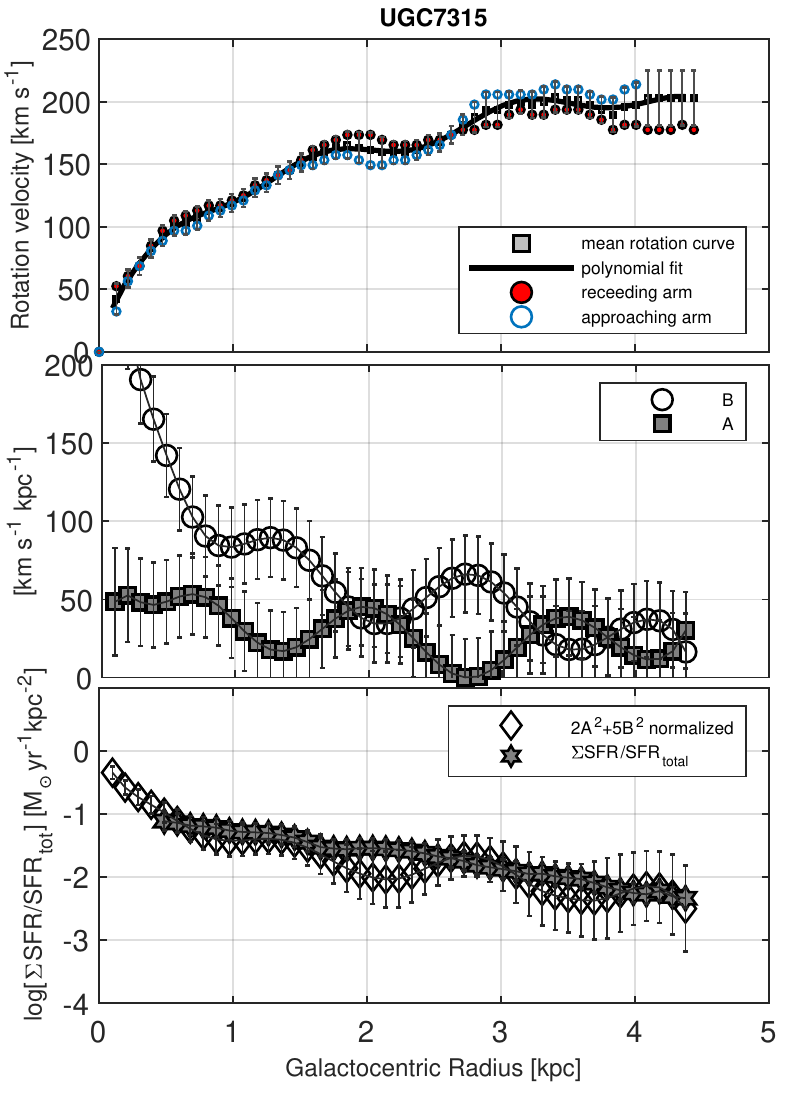}
\end{tabular}
	\end{figure*}
	
\label{Figure:5}






\label{lastpage}

\end{document}